\documentclass[a4paper,12pt,sort&compress,3p]{elsarticle}

\usepackage[T1]{fontenc} % Use modern font encodings 
\usepackage[utf8]{inputenc}
\usepackage{lmodern}

\usepackage{subfigure}
\usepackage{siunitx}

\usepackage{url}
\usepackage{hyperref}
\hypersetup{colorlinks=true}
\usepackage[english]{babel}

\usepackage{amsfonts}
\usepackage{amsmath}
\usepackage{amssymb}

\newcommand{\R}{\mathbb{R}}
\newcommand{\set}[1]{\left\{#1\right\}}

\newcommand{\refe}{\mathrm{ref}}
\newcommand{\simu}{\mathrm{sim}}
\newcommand{\treach}{t_{\mathrm{reach}}}
\newcommand{\tcheck}{t_{\mathrm{check}}}
\newcommand{\tgr}{t_{\mathrm{G-R}}}
\newcommand{\tfv}{t_{\mathrm{F-V}}}
\newcommand{\tsim}{t_\simu}
\newcommand{\obs}{\mathcal{O}}
\newcommand{\tol}{{\rm TOL}}
\newcommand{\prob}{\mathbb{P}}
\newcommand{\gparrep}{Gen.~ParRep~}
\newcommand{\argmin}{\arg\!\min}

\journal{Computer Physics Communications}

\newcounter{bla}

\begin{document}

\begin{frontmatter}



\title{gen.parRep: a first implementation of the Generalized Parallel Replica dynamics
for the long time simulation of metastable biochemical systems}


\author[enpc,inria]{Florent H\'{e}din}
\ead{florent.hedin@enpc.fr}

\author[enpc,inria]{Tony Leli\`{e}vre}
\ead{tony.lelievre@enpc.fr}

\address[enpc]{CERMICS, \'{E}cole des Ponts--ParisTech,
6 et 8 avenue Blaise Pascal,
Cit\'{e} Descartes -- Champs-sur-Marne,
F-77455 Marne-la-Vall\'{e}e Cedex 2, France}

\address[inria]{Inria Centre de Recherche Paris Rocquencourt, 2 rue Simone Iff,
F-75589 Paris, France}

\begin{abstract}
Metastability is one of the major encountered obstacle when performing long molecular dynamics simulations,
and many methods were developed to address this challenge.
The ``Parallel Replica''(ParRep) dynamics is known for allowing to simulate very long
trajectories of metastable Langevin dynamics in the materials science community,
but it relies on assumptions that can hardly be transposed to the world of biochemical simulations.
The later developed ``Generalized ParRep'' variant solves those issues, but it was not applied to
significant systems of interest so far.\\

In this article, we present the program \emph{gen.parRep}, the first publicly available
implementation of the Generalized Parallel 
Replica method
(BSD 3-Clause license), targeting
frequently encountered metastable biochemical systems, such as conformational equilibria or 
dissociation of protein--ligand complexes.
It will be shown that the resulting C++ implementation exhibits a strong linear scalability, providing up to 70\% of the maximum possible 
speedup on several hundreds of CPUs.\\
\end{abstract}

\begin{keyword}
Molecular simulations; Langevin dynamics; Quasi-stationary distributions; Metastability; Protein–Ligand dissociation; High Performance 
Computing
\end{keyword}

\end{frontmatter}

\clearpage

\begin{small}
\noindent
{\em Program Title:}                   \\
gen.parRep\\
{\em Licensing provisions:}            \\
BSD 3-clause\\
{\em Programming language:}            \\
C++ (mostly), C and Lua\\
%{\em Supplementary material:}                                 \\
%  % Fill in if necessary, otherwise leave out.
%{\em Journal reference of previous version:}                  \\
%  %Only required for a New Version summary, otherwise leave out.
%{\em Does the new version supersede the previous version?:}   \\
%  %Only required for a New Version summary, otherwise leave out.
%{\em Reasons for the new version:}\\
%  %Only required for a New Version summary, otherwise leave out.
%{\em Summary of revisions:}*\\
%  %Only required for a New Version summary, otherwise leave out.
{\em Nature of problem:}\\
Molecular dynamics simulations of chemical and biological systems usually encounter the problem of 
metastability, because of the timescale separation between the time discretization step used for dynamics
and the usual mean time between conformational changes. The use of Accelerated dynamics [1] methods is
usually necessary in order to address this challenge.\\
{\em Solution method:}\\
The Generalized Parallel Replica method [2] accelerates the exit from metastable states, providing a linear speedup of $N$, $N$ being the 
number of replicas of the system running in parallel. This C++ implementation, the first available so far, exhibits a strong linear scaling 
on hundreds of CPUs, therefore ready for production studies on High Performance Computing (HPC) machines.\\
{\em Additional comments:}\\
Git repository: 
\href{https://gitlab.inria.fr/parallel-replica/gen.parRep}{https://gitlab.inria.fr/parallel-replica/gen.parRep}
\\

\end{small}

\clearpage

\section{Introduction}
\label{sec:intro}

Molecular dynamics (MD) simulations are nowadays of a common use for simulating large and
complex biological or chemical systems \cite{review_md_bio}:
the continuous increase of the available computing power, together with the development of
stable and accurate deterministic or stochastic sampling strategies, made possible the emergence of computer based,
\emph{in silico} drug design strategies \cite{review_md_drug_design,lounnas_current_2013,sliwoski_computational_2014}.
However, a commonly encountered obstacle while performing MD simulation is the timescale separation between the fastest
conformational changes --- usually vibrations occurring at the femtoseconds (fs) level --- and the slowest one,
occurring from the nanosecond (ns) to second (or more) timescale; one may use various
coarse grained \cite{cg_bio_1,cg_bio_2}
approaches in order to reach such large simulation time, however this usually implies to sacrifice the
accurate description
of fast processes, such as non-bonded donor--acceptor interactions, playing a key role in biological 
interactions \cite{non_bonded_bio1,non_bonded_bio2}.
The existence of \emph{metastable} regions in the configurational space, separated by high potential energy
or entropy barriers, is 
the main origin of this timescale separation, and the simulation time required for observing a transition
from such a region to another one
can quickly become intractable by the use of direct numerical simulations. 

A large amount of methods were developed to address the challenge of metastability in MD simulations.
When it is assumed that both the starting and the ending metastable regions
(let us denote them by $A$ and $B$) are known, one can consider that most of the methods fall within one of the two 
following categories:
\emph{local search methods} start from an initial guess path connecting $A$ and $B$, and will optimize it until
convergence to an optimal path, for example characterized by a minimal potential or free energy profile:
the nudged elastic band method \cite{neb_method},
the string method \cite{string_method},
the max flux approach \cite{max_flux_method},
the weighted ensemble methods \cite{Huber1996Jan,WESTPA_2018Dec},
or the transition path sampling method \cite{tps_method} (which is actually a path sampling method starting from the initial guess, but not 
an optimization method);
the second category consists in \emph{global search methods} where the ensemble of all the possible
paths between $A$ and $B$
is sampled without any initial guess, and it includes
adaptive multilevel splitting 
methods \cite{Cerou_ams_2007,Cerou_ams_2011,brehier_ams_2015,teo_ams_2016},
transition interface sampling \cite{tis_method},
forward flux sampling \cite{fflux_method}
or milestoning techniques \cite{Faradjian_milestoning_2004,Maragliano_milestoning_2009,Schutte_milestoning_2011,
bellorivas_milestoning_2016,aristoff_milestoning_2016,Grazioli2018Aug_part1,Grazioli2018Aug_part2}.

A. F. Voter and coworkers also proposed another class of methods, the \emph{Accelerated Dynamics 
methods} \cite{Voter_accMD_1998,Perez_accMD_2009,Lelievre_accMD_2015,Lelievre_stoltz_2016,Lelievre_accMD_2018}:
the Parallel Replica (ParRep) method (and the derived ParSplice
algorithm) \cite{voter_parrep_1998,kum_parrep_2004,perez_parsplice_2016,perez_parsplice_code_2017},
the hyperdynamics method \cite{voter_hyper_1997,voter_hyper2_1997},
and the temperature accelerated dynamics method \cite{Sorensen_tad_2000}.
They all
% have in common to be derived from
rely on 
the Transition State theory and kinetic Monte Carlo models,
and the aim of these algorithms is to efficiently generate the succession of jumps
between metastable regions in a statistically consistent way compared to the reference Langevin dynamics.

The ParRep method was later formalized \cite{bris_parrep_2011}, and it was shown that
the notion of quasi-stationary distribution (QSD) \cite{ferrari_qsd_1995,doorn_qsd_2011} is the mathematical
foundation at its heart: this revealed one of the possible
weaknesses of the algorithm, where it is assumed that the (user defined) time required for converging to the
QSD is the same for all the metastable regions.
While this assumption may be reasonable for materials science, this cannot be transposed
to the chemical configurational space,
where the large variety of possible interactions and steric exclusions usually results in a rough
energy landscape, characterized by both 
an extremely large number of energy minima, and the presence of super basins of attraction
(usually referred to as ``funnels'').
The Generalized Parallel Replica (\gparrep) \cite{binder_genparrep_2015} method addresses this issue by
estimating during the simulation if convergence to the QSD is obtained; however, while this can possibly 
extend the range of application of ParRep to any biochemical system which can be studied via MD,
no implementation has been designed and released so far.
 
This article describes the first publicly available implementation of \gparrep,
specially targeting metastable biochemical systems.
After a description of the methods in Section \ref{sec:methods},
the novelty of the software implementation is detailed in Section \ref{sec:softwareImpl};
two study cases are later investigated in Section \ref{sec:results}, the conformational equilibrium 
of the alanine dipeptide (subsection~\ref{subsec:ala2}), and the dissociation of the
protein--ligand complex FKBP--DMSO (subsection~\ref{subsec:fkbp}).
It will be shown in both cases that the \gparrep algorithm can be used for accurately
sampling the state-to-state dynamics, and in particular the state-to-state transition times;
furthermore evidences that the 
software exhibits a strong linear scaling will be reported: when
running over hundreds of CPUs, one gets speedups of up to 70 \% of the maximum possible linear speedup.\\

%%%%%%%%%%%%%%%%%%%%%%%%%%%%%%%%%%%%%%%%%%%%%%%%%%%%%%%%%%%%%%%%%%%%%%%%%%%%
\section{Methods}
\label{sec:methods}

\subsection{Langevin dynamics}
\label{subsec:langevin}

Let us consider a stochastic process $X_t = (q_t,p_t)_{t \ge 0} \in \R^{d \times d}$ ($\R^{d \times d}$ representing 
the \emph{phase space}), where $q$ and $p$ denote the positions and momenta of the $d/3$ particles at time $t$.
The stochastic process $X_t$ follows the Langevin dynamics:
\begin{equation}
\begin{cases}
dq_t = M^{-1} p_t \mathop{dt} \\
dp_t = - \nabla V(q_t) \mathop{dt} -\gamma M^{-1} p_t \mathop{dt} + \sqrt{2\gamma\beta^{-1}} \mathop{dW_t}
\end{cases} 
\label{eqn:langevinDynamics}
\end{equation}
where 
$\beta = \frac{1}{k_B T}$ is the inverse temperature, $M$ is the mass matrix,
$V : \R^d \rightarrow \R $ is a function associating to a given configuration $q$ a potential energy $V(q)$,
$\gamma > 0$ is the damping parameter, and $W_t$ a $d$-dimensional Brownian motion.

The Langevin dynamics on the $d$-dimensional potential energy surface $V$ is likely to consist in a succession of 
``entry then exit'' events from wells (or groups of wells) progressively discovered by the process $X_t$,
and one can expect that the 
time spent within a well before it hops to another one will be far more large that the discretization timestep $dt$:
it is therefore necessary to design an alternative approach to the computationally expansive direct
simulation in order to address this problem of \emph{metastability}.

%-----------------------------

\subsection{States and metastability}
\label{subsec:states}

Let us introduce the ensemble of metastable states $\mathcal S=\{S_1, ... , S_n\}$. These are typically defined in terms of positions only 
(and not velocities). In the original ParRep algorithm \cite{voter_parrep_1998,perez_parsplice_2016}, these states are defined as the 
basins of attraction of the local minima of $V$ for the gradient descent $\dot{q} = - \nabla V(q)$: this leads to a partition of the state 
space.
One important output of the mathematical analysis performed in Ref. \cite{bris_parrep_2011} is that
(i) the metastable states can be defined arbitrarily, the only prerequisite being that for most of the visits in one of those, the exit 
time will be much larger than the convergence time to the local equilibrium within the state (the so-called Quasi Stationary Distribution), 
and
(ii) the algorithm can be applied even if these metastable states do not define a partition of the state space:
in this work we propose to define them as disjoint subsets, using \emph{collective variables} or \emph{reaction 
coordinates} \cite{colvars_2013,plumed2_2014},
modeled a priori in order to correspond to a few given metastable conformations of the molecular 
system of interest; the topological definition of the states will be discussed for each 
system of interest in the Section \ref{sec:results}.\\

Let $\Omega \in \mathcal{S}$ be a given state: we define
\[
\tau = \inf\set{t\geq 0\mid X_t \notin \Omega}
\]
to be the first exit time from $\Omega$ (for a given initial condition $X_0 \in \Omega$), and
\[ 
X_\tau \in \partial \Omega
\]
to be the corresponding exit configuration (first hitting point on the boundary $\partial\Omega$):
the goal of the various Parallel Replica (ParRep) \cite{voter_parrep_1998,binder_genparrep_2015,perez_parsplice_2016} based methods
(but also of other accelerated dynamics methods)
is to sample efficiently the values $(\tau,X_\tau)$ from the unknown exit distribution associated to each state $\Omega$.

%-----------------------------

\subsection{Quasi-Stationary Distribution (QSD)}
\label{subsec:qsd}

Recent mathematical analyses showed \cite{bris_parrep_2011} that
the \emph{quasi-stationary distribution} (QSD) \cite{ferrari_qsd_1995,doorn_qsd_2011} is an essential
ingredient of the above mentioned accelerated dynamics methods.
Let~$\nu$ be a probability measure with support in $\Omega$: $\nu$ is a QSD if and only if, for any $A \subset 
\Omega$ and $t \ge 0$:
\[
\nu(A) = \prob^{\nu} \left[ X_t \in A\mid t <\tau \right]
\]
where $\prob^{\nu}$ indicates that the initial condition $X_0$ is distributed according to $\nu$.
This means that $\nu$ is a QSD if, for all $t$, when $X_0$ is distributed according to $\nu$, the law
of $X_t$ conditionally to the fact that $(X_s)_{0 \le s \le t}$ remains in the state $\Omega$ is still $\nu$.

The QSD satisfies the following properties which will be of critical importance (see Refs.\cite{bris_parrep_2011,binder_genparrep_2015} for 
detailed proofs):
\begin{enumerate}
\item Existence and uniqueness of $\nu$: the QSD is the unique long time limit ($t \rightarrow + \infty$)
of the distribution of $X_t$, conditioned to starting and staying in $\Omega$ up to time $t$;
\item if $X_0$ is distributed according to the QSD $\nu$, 
then the first sampled exit time $\tau$ is independent of the first sampled exit configuration $X_\tau$;
\item if $X_0$ is distributed according to the QSD $\nu$, sampled values of the first exit time $\tau$ are exponentially distributed: 
$\prob^{\nu}(\tau>t) = e^{-\lambda t}$, where $\lambda = \frac{1}{\mathbb{E}^{\nu}(\tau)}$.
\end{enumerate}

%----------------------------------------------------
\subsection{The Generalized ParRep method}
\label{subsec:genParRep}

Having introduced the concepts of states and QSD, it is now possible to detail the Generalized Parallel Replica 
\cite{binder_genparrep_2015} 
(\gparrep) method. 
In the following, it is assumed that different metastable states $\mathcal{S} = \{S_1,...,S_n\}$ are defined, either by 
partitioning the whole configuration space, or by defining disjoint subsets of $\R^d$, and $\Omega$ denotes any member of $\mathcal{S}$. 
It is also assumed that at least
$N$ CPU cores are available in order to propagate simultaneously $N$ replicas of the system in parallel.

As stated above, the aim of accelerated dynamics methods is to quickly sample values of $(\tau,X_\tau)$ (respectively the first exit
time from a metastable $\Omega \in \mathcal{S}$ and the first hitting point on the boundary $\partial\Omega$): in the case of ParRep 
methods, detailed information about how the process evolves within each state $\Omega$ is discarded, and in return exit events can be 
generated $N$ times faster (a linear speedup is achieved), which is of particular interest when considering computations performed on High 
Performance Computing (HPC) machines, where thousands of CPUs can be used at once by a single simulation.
\\

\begin{figure}[h!]
\centering
\includegraphics[width=0.85\linewidth]{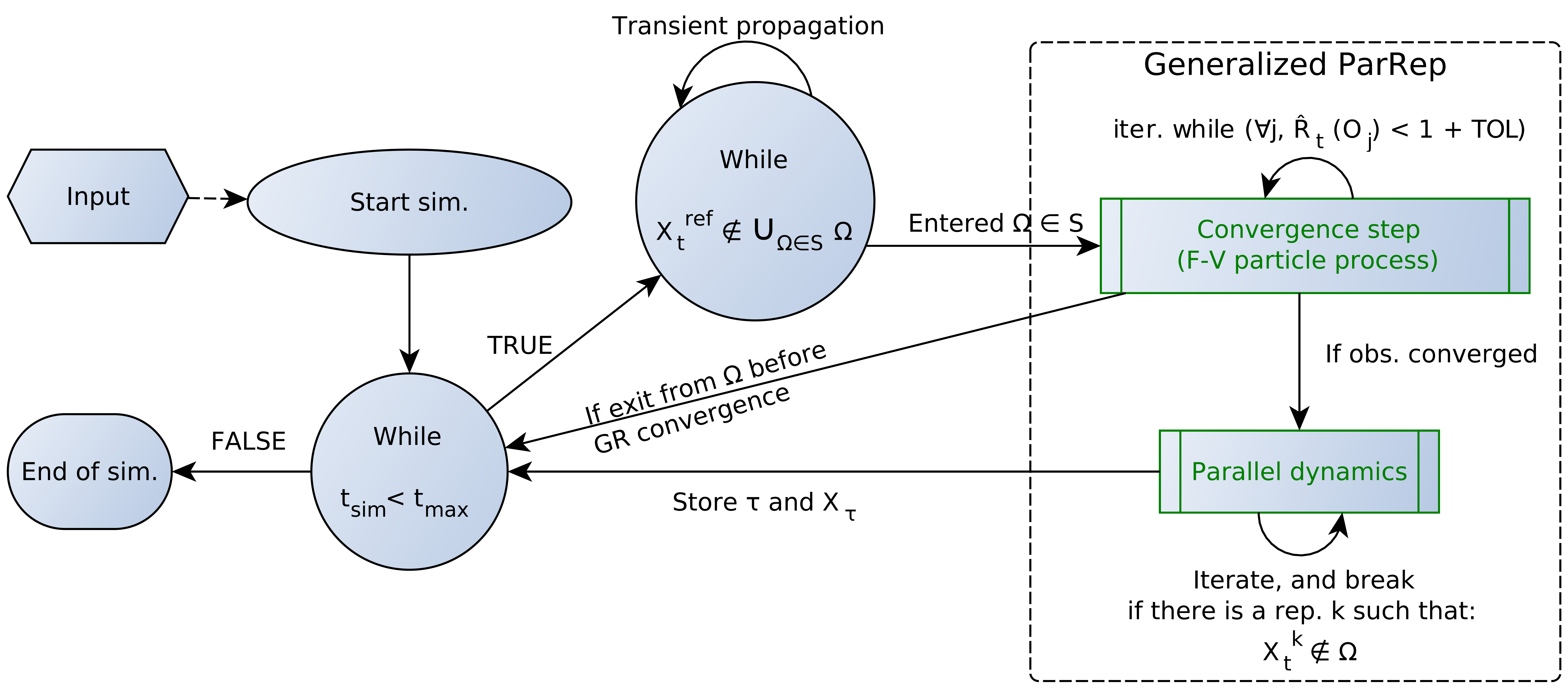}
\caption{Diagram view of the generalized ParRep \cite{binder_genparrep_2015} algorithm.
After setup, the first step (Transient propagation) is to iterate the reference walker until $X_t$ enters a defined state $\Omega$
(if the states define a partition of the configuration space, this first part of the algorithm is not required); then
the \gparrep procedure (right frame) is executed, starting with the Convergence step, until either:
(i) the reference walker exits before convergence of the G-R statistics is observed, or
(ii) convergence of the G-R observables is obtained before the reference walker exits.
In the later case simulation proceeds to the Parallel dynamics step, until an exit event from $\Omega$ is observed, generating a sample
of $(\tau,X_\tau)$.
After the parallel phase (or if $X_t$ exited $\Omega$ before convergence), the reference walker performs once again the Transient 
propagation procedure, until entering a valid 
state~$\Omega$, and the \gparrep procedure is iterated once again.
This is repeated until the total simulation time $\tsim$ reaches a user defined value $t_{\rm max}$, where the program stops.
The two frames colored in green are parts of the algorithm fully exploiting the $N$ available CPU cores.
}
\label{fig:algorithm_gen_parrep}
\end{figure}

In the following, let $\tsim \ge 0$ be the simulation clock, corresponding to the physical time (i.e. a multiple of the time step $dt$),
and let $X^\refe_{\tsim}$ be the configuration of the system at time $\tsim$ (where $\refe$ indicates the \emph{reference walker}, the 
first replica).
The method is implemented as a three steps procedure, repeated as the process diffuses from one state to
another, until a total simulation time $t_{\rm max}$ is reached (see Figure~\ref{fig:algorithm_gen_parrep} for a diagram representation):

\begin{enumerate}
\item Transient propagation: if the set $\mathcal{S}$ is not a partition of the whole configuration space, it might be that 
$X^\refe_{\tsim}$ is outside of any known state:
therefore the process has to be propagated for a time $\treach$ until it reaches a metastable 
state $\Omega$ (note that $\treach$ is expected to be much smaller than the typical exit times from the states in $\mathcal{S}$, at least 
if the states definitions encompass accurately the metastable domains). After 
this step, the simulation time is updated as $\tsim \leftarrow \tsim + \treach$.

\item Convergence step: a Fleming-Viot (F-V) particle process is launched to estimate the convergence time to the QSD.
If the reference walker leaves $\Omega$ before the convergence time to the QSD, one goes back to step 1.
If not, one proceeds to the Parallel dynamics step.

\item Parallel dynamics step : $N$ replica are propagated independently in parallel, until one exits the state $\Omega$. The 
corresponding exit time $\tau$ is calculated (more details below) 
and is saved together with the exit configuration $X_\tau$; then the program proceeds to a new Transient propagation.
\end{enumerate}

In terms of wall-clock time, the computaional gain of this algorithm compared to a direct numerical simulation comes from the parallel 
dynamics step, which allows to generate a sample of the exit event $(\tau,X_\tau)$ in a wall-clock time $N$ times smaller than for the 
direct numerical simulation.\\

In the following the Convergence step and Parallel dynamics step will be detailed.\\

%----------------------------------------------------

\subsubsection{Convergence step: Fleming-Viot process and Gelman-Rubin convergence diagnostic}

The Fleming-Viot (F-V) process \cite{FV_1979,ferrari_fv_2011} is a branching and interacting particle process,
used for simulating the law of the random variable $X_t$ conditioned to $\{ \tau > t \}$.
As a consequence, an estimate of $\tfv$ --- the F-V convergence time --- can be obtained by assessing the convergence to a stationary state 
of the F-V process, and when this convergence is observed, one obtains samples (approximately) distributed according to 
the QSD.
For a detailed description with illustrations, we refer to the dedicated section from 
Ref. \cite{binder_genparrep_2015}.

Let us first consider $N$ i.i.d. initial conditions $X_0^k$ ($k \in \{1, \ldots,N\}$); the procedure summarizes as 
follows:
a \emph{reference walker} $X^\refe_{t}$ (namely the replica numbered $k=1$) explores $\Omega$ driven by the Langevin 
Equation~(\ref{eqn:langevinDynamics}): 
at the same time the other replicas (the \emph{F-V workers})
perform the following tasks:
\begin{enumerate}
\item the F-V workers evolve independently according to Equation~(\ref{eqn:langevinDynamics}) within $\Omega$, each of 
them regularly collecting the instant values of several \emph{observables}; until one of them, e.g. 
$X^i$, exits;
\item the process $i$ that exits is discarded, and replaced by a copy of one of the other F-V workers 
(\emph{survivors}), randomly drawn with uniform probability among the survivors: this is called a \emph{F-V branching};
\item the survivors and the newly branched processes evolve and collect values, going back to 1., 
until convergence is reached for each observable (convergence will be 
defined below using the Gelman-Rubin diagnostic).
\end{enumerate}
However, if at any moment the reference walker $X^\refe$ leaves $\Omega$ before the F-V process has converged,
all the F-V walkers replicas are killed, and
a new Transient propagation is initiated.
\\

The observables are properties of interest which are expected to characterize the convergence to equilibrium of the F-V particle process 
within each state $\Omega$: they can have a physical meaning (e.g. based on the potential $V$, or the momenta $p$), or be any type 
of distance/topological measure (for instance derived from the collective variables used for designing the sets in $\mathcal{S}$).\\

The convergence of the observables is assessed using the Gelman-Rubin (G-R) statistics \cite{gr_1992,gr_1998}:
let $\obs : \Omega \to \R$ be some observable, and let
\begin{align}
\begin{split}
\bar{\obs}^k_{t} &\equiv t^{-1} \int_0^t \obs(X_s^k)~ds \\
\bar{\obs}_{t}   &\equiv \frac{1}{N} \sum_{k=1}^N \bar{\obs}^k_{t} = \frac{1}{N} \sum_{k=1}^N t^{-1} \int_0^t \obs(X_s^k)~ds
\end{split}
\label{eqn:GR_def_Obs}
\end{align}
be the average of an observable along each trajectory ($\bar{\obs}^k_{t}$) and the average
of the observable along all trajectories $\bar{\obs}_{t}$. The statistic of interest for the observable $\obs$ is defined by:
\begin{equation}
\hat{R}_{t}(\obs) = \frac{\frac{1}{N}\sum_{k=1}^N t^{-1} \int_0^t(\obs(X_s^k) - \bar{\obs}_{t})^2 ds }{\frac{1}{N}
\sum_{k=1}^N t^{-1} \int_0^t (\obs(X_s^k) -\bar{\obs}_{t}^k )^2 ds }
\label{eqn:GR_def_ratio}
\end{equation}
Note that $\hat R_t(\obs) \geq 1$, and as the F-V workers' trajectories
explore $\Omega$, $\hat{R}_t(\obs)$ converges to $1$ as $t$ goes to infinity.

The time required for the F-V particle process to converge is denoted by $\tfv$ and defined by:
\begin{equation}
\tfv = \inf \set{t\geq 0\mid \hat{R}_t(\obs_j) < 1 +  \tol,~\forall j}
\label{eqn:GR_tfv_formula}
\end{equation}
i.e. it is the time required for obtaining a ratio $\hat{R}_{t}(\obs)$ less than $1 + \tol$ for each of the observable $\obs$
(where $\tol > 0$ is a user defined stopping criterion).

After a successful Convergence step, the simulation clock is updated as follows:
\begin{equation*}
\tsim \leftarrow \tsim + \tfv
\end{equation*}
and one proceeds to the Parallel dynamics step; in case the reference walker left $\Omega$ before the convergence time $\tfv$, the 
simulation clock time is updated as follows:
\begin{equation*}
\tsim \leftarrow \tsim + t_{\refe}
\end{equation*}
where $t_{\refe}$ is the amount of simulation time the reference walker spent within $\Omega$ before an exit event was observed, and one 
proceeds to a new Transient propagation step.

Note that because of the small value of the timestep $dt$, usually chosen between $0.5$ and $2~\rm fs$, one does 
not expect to observe large fluctuations of the observables between two consecutive times $t$ and $t + dt$: it therefore makes sense to 
accumulate the values of the observables less frequently, say with a period $\tgr$, satisfying $dt < \tgr \ll \tfv$.

Likewise, the test to check whether an exit from $\Omega$ occurred is only performed with period $\tcheck$, with typically 
$\tgr < \tcheck \ll \tfv$.

\subsubsection{Parallel dynamics step}

The $N$ samples obtained after the Convergence step are used as initial conditions; then the $N$ replicas are propagated following 
Equation~(\ref{eqn:langevinDynamics}) with independent driving Brownian motions.

Let $t_{\rm para} = 0$ be the simulation time spent in the Parallel dynamic step, until the first exit event is observed;
let $\tcheck$ be a simulation time interval (multiple of $dt$) at which one tests if an exit event occurred, and let $M$ counts how many 
times this test was performed before an exit event occurred;
finally let
\begin{equation*}
k = \min \argmin_{n \in \{1,...,N\}} t_{\rm para}^n
\end{equation*}
be the index of the first replica for which an exit event occurred:
then it was shown \cite{aristoff_discrete_2014} that the exit time $\tau$ can be sampled as:
\begin{equation}
\tau = \left[ N(M-1) + k \right] \tcheck .
\label{eqn:ParRep_CorrectedDiscreteTauEscape}
\end{equation}

The simulation clock is updated as:
\begin{equation}
\tsim \leftarrow \tsim + \tau .
\label{eqn:ParRep_ClockUpdate}
\end{equation}
A new Transient propagation can therefore be initiated, using as new initial condition the exit point $X_\tau^k$ of the first replica which 
exited.\\

\subsubsection{Differences with the original ParRep algorithm}

The \gparrep algorithms differs from the original ParRep algorithm (as described in Refs. \cite{voter_parrep_1998,kum_parrep_2004}) on 
several points:

\begin{itemize}
\item The original ParRep algorithm has originally been introduced on a partitioned configuration space, usually defining
states as the basins of attraction of the local minima of the potential energy function, thus implying regular gradient descent on $V$.
This makes the state identification simple and unambiguous for systems characterized by a smooth potential energy landscape
where minima are separated by high energy barriers;
however biochemical systems are usually characterized by rough and funneled energy landscapes, where conformation changes usually 
involve numerous transitions over local minima separated by low energy barrier.

\item The original ParRep implementations require the user to define two parameters, the decorrelation time $t_{\rm corr}$ and the 
dephasing time $t_{\rm phase}$.
The decorrelation time $t_{\rm corr}$ is used to assess the convergence to the QSD for the reference walker: if it stays in a state $\Omega$
for a time $t_{\rm corr}$ it is assumed to be distributed according to the QSD.
Likewise, the dephasing time $t_{\rm phase}$ is used to sample the QSD before the Parallel dynamics step starts: in the 
so-called dephasing step, each of the $N$ replica is propagated within the state $\Omega$, and its end point is kept as a sample of the QSD 
if it stayed within the state $\Omega$ for a time $t_{\rm phase}$.
Once again, this approach appears hardly compatible with biochemical systems, as it is impossible to define ubiquitous values 
of $t_{\rm corr}$ and $t_{\rm phase}$ appropriate for all the possible local minima and all initial conditions within the states.

\end{itemize}

Those two limitations are addressed by the implementation of the \gparrep algorithm described in this article: while permitted, partition 
of the configuration space is not enforced, and the user has total control on how to define the states; this allows for instance to merge 
multiple local minima together in order to define a metastable state accurately englobing a funnel of the PES.

Furthermore the use of the F-V particle process during the Convergence step releases the user from providing a priori estimates of the time 
required for converging to the QSD, as $\tfv$ is estimated on the fly based on the convergence of the observables, the only requirements 
being to provide meaningful observables and a tolerance level.

%%%%%%%%%%%%%%%%%%%%%%%%%%%%%%%%%%%%%%%%%%%%%%%%%%%%%%%%%%%%%%%%%%%%%%%%%%%%
\section{Software implementation}
\label{sec:softwareImpl}

In the following section \ref{sec:results} we will present results obtained with our current implementation of the 
Generalized ParRep algorithm: it consists in a newly written C++ program, \emph{gen.parRep}, available
free of charge (see \href{https://gitlab.inria.fr/parallel-replica/gen.parRep}{https://gitlab.inria.fr/parallel-replica/gen.parRep})
and released under an open-source BSD 3-clause licensing. We aimed at providing an easy to use, 
versatile and performance oriented implementation, focusing on the study of metastability encountered when studying 
chemical and biochemical systems. Note that while the original ParRep method is also implemented and available in our new software, 
we will not present any result for it, as we focus on the novelty of \gparrep.

In the following paragraphs, the critical requirements for developing such a code are detailed, together with details 
on the technical solutions adopted in order to address them.\\

\subsection{Distributed computing capabilities}

The replica-based approach of the ParRep algorithms naturally suggests that the parallelization is achieved by
using a distributed computing approach: an obvious choice nowadays is to use the Message Passing Interface 
(MPI) \cite{mpi_1994} standardized protocol, for which various high performance computing (HPC) implementations are 
available \cite{openmpi_2004,mpich2_2002}.

Each of the $N$ replica corresponds to a \emph{MPI task}: each task will use $P$ CPU cores, $P$ being 
at least $1$ and at most all the cores available on a given machine (a \emph{MPI node}).
Therefore each computing node will execute $1$ or more replicas, each performing the dynamics on $P$ cores.

Regularly, \emph{messages} of arbitrary size are exchanged between the replicas, which can be classified in two 
categories: 
\begin{itemize}
\item \emph{point-to-point} communications involve two replicas and are usually inexpensive as long as the amount of 
data sent remains relatively small: one example is the branching and cloning operation of the F-V algorithm, where an 
exiting F-V worker will copy the $X_t = (q_t,p_t)$ configurations plus the history of all the $\obs$ observables from 
another F-V worker.
\item \emph{collective} communications involve the full ensemble of the $N$ replicas and are likely to be time 
consuming, and are therefore used with care: they include \emph{barriers} for keeping the replicas synchronized
and \emph{broadcasting} operations where a replica sends its configuration $X_t = (q_t,p_t)$ to the $(N-1)$ others (for 
example to be used as an initial condition for the next F-V iteration).
\end{itemize}
Furthermore, communications can either be \emph{blocking} or \emph{non-blocking}, the later allowing the developer to interleave 
communications and computations in order to hide latency. To provide an efficient \gparrep implementation, the use of 
barriers and collective communications have been reduced to the minimum possible, and non-blocking variants of those were used whenever 
possible.\\

\subsection{MD engine}

One requires an efficient Molecular Dynamics (MD) engine, capable of performing the dynamics of Equation~(\ref{eqn:langevinDynamics}): the 
minimal requirement is to have access to one code block which, when executed, will 
realize one or more 
discretization steps of size $dt$, and which internally takes care of the evaluation of the potential $V(q)$ and its 
gradient (usually analytically calculated). A read and write access to the internal configuration $X_t = 
(q_t,p_t)$ of each replica is also required for performing the exchanges.

In order to study large systems, one also expects: full support of commonly used 
force-fields, availability of modern optimizations such as 
the Particle Mesh Ewald \cite{Darden_pme_1993}, Reaction Field \cite{Lee_RF_1992,Tironi_RF_1995}, or Cell-Linked Lists
\cite{Mattson_celllink_1999,Yao_celllink_2004,Heinz_celllink_2004,Gonnet_celllink_2007} methods, for an efficient 
evaluation of non-bonded interactions. As mentioned in the previous paragraph one can decide to provide $P \ge 1$ CPU 
cores to each of the $N$ replica, therefore a shared memory parallelization capability for the MD engine is encouraged.

For the current implementation it was decided to use the OpenMM 7
library; \cite{omm7_2017} OpenMM is a high performance, free of charge 
and open source toolkit for performing molecular simulations, which can be used either as a software library on which 
to build a program, or directly as an application (via python scripting): the later is used for preparing the molecular 
systems before simulation, accepting force-field and configuration files from various origins 
(CHARMM \cite{charmm_2009}, 
AMBER \cite{amber_2017}, GROMACS \cite{gromacs_2015}, NAMD \cite{namd_2005},...), while the library mode provides a direct 
and simplified access to the MD engine from the C++ application.\\

\subsection{Definition of the states $\mathcal{S}$}

While technical aspects as parallelization and efficiency of the MD engine are important, the Generalized ParRep
critically relies on an efficient definition of the set of states $\mathcal{S}$. As stated in subsection~\ref{subsec:states}
this implementation focuses on applications where the states are a priori defined using either atomic coordinates 
or more elaborated collective variables: it is thus necessary to provide a way to define the states \emph{online}, 
e.g. using a scripting language interfaced with the core C++ methods in order to have access to atomic properties.

\begin{figure}[h!]
\centering
\includegraphics[width=\linewidth]{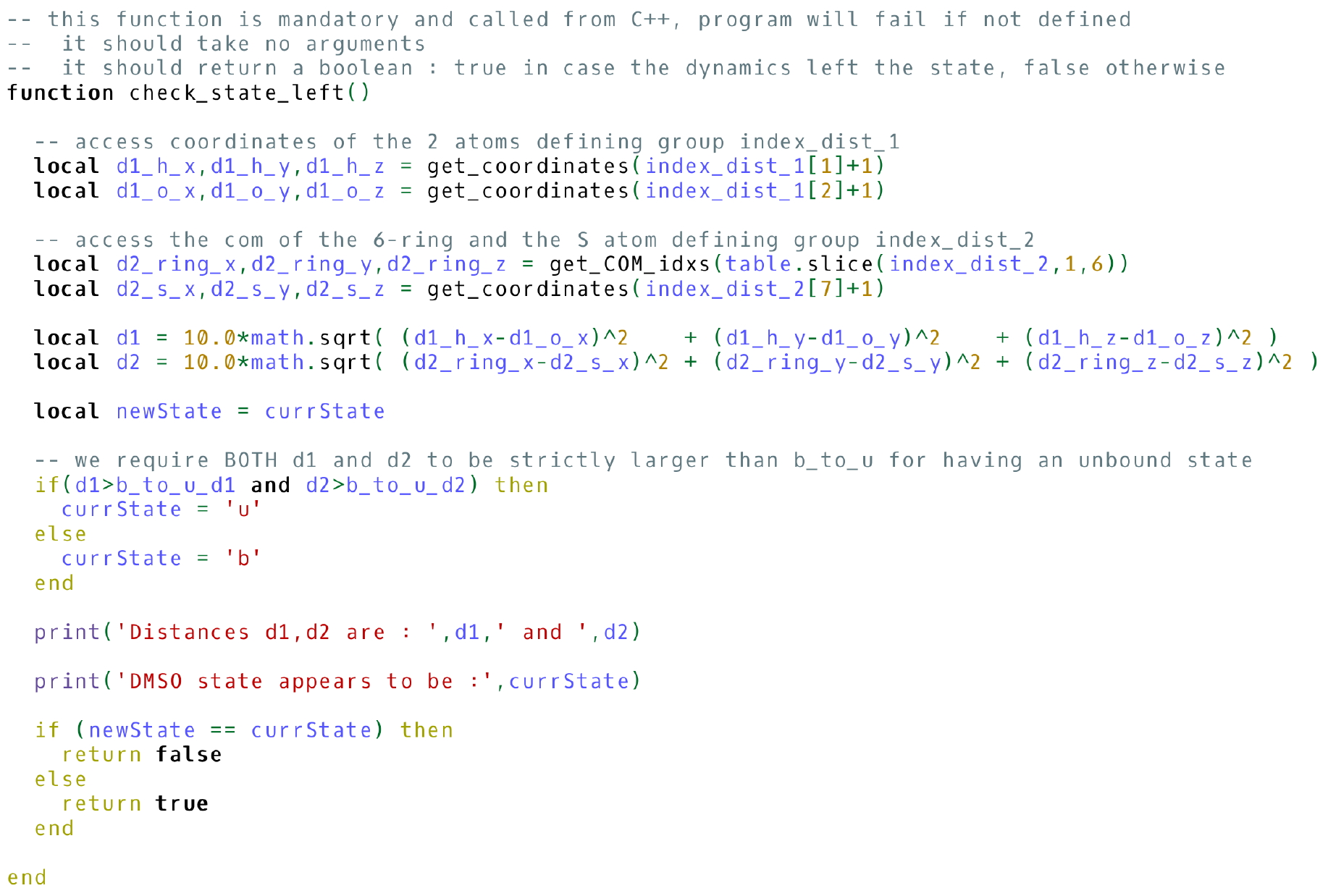}
\caption{
Example of a Lua function written by the user within the input file (corresponding to the validation system presented in subsection 
\ref{subsec:fkbp} below), and called from the C++ program.
This function is called every time the algorithm checks whether an exit event from the current metastable state
$\Omega$ happened, either during the Convergence Step or the Parallel dynamics step.
The variables \texttt{index\_dist\_1} and \texttt{index\_dist\_2} are simply tables of atomic indices defined earlier in the input file,
and indices used in the state definition (see Figure \ref{fig:FKBP_cavity_dists} (b) below).
Functions \texttt{get\_coordinates(sele)} and \texttt{get\_COM\_idxs(sele)} are bindings to the C++ code which respectively retrieve
atomic coordinates, and calculate the center of mass, for a given set of atomic indices \texttt{sele}.
This versatile procedure gives to the user a lot of flexibility for: (i) defining the metastable states and (ii) detecting exits, 
as it does not require any modification of the compiled code.
\label{fig:lua_code_example}
}
\end{figure}

It was decided to use the Lua \cite{lua_1996} language: it is a fast, lightweight, easy to learn,
embeddable and dynamically typed scripting language. 
The user input required for running the ParRep algorithms is written to an input Lua file,
together with all the code and variables for
(i) defining the states,
(ii) checking if an exit event is observed, and
(iii) monitor the G-R statistics.
The Sol2 \cite{sol2} library (embedded within the C++ program's source code) takes care of parsing the input 
file at initialization, and it dynamically maps the user-defined code to C++ functions. The core code is therefore 
\emph{state agnostic} as it never exactly knows how a state has been defined: indeed the whole implementation will only call the following:
(i) a function returning a true/false boolean value indicating if $(X_t \notin \mathcal{S})$ (always true in case of a partitioned 
configuration space, possibly false otherwise);
(ii) another function returning a true/false boolean value indicating if $(X_t \notin \Omega)$ where $\Omega$ is the last visited state, 
being called every time it is required to check if an exit event occurred;
(iii) and a few functions (one per user defined observable) monitoring the G-R observables $\obs$ returning a real value to be accumulated 
and used in 
Equations (\ref{eqn:GR_def_Obs}), (\ref{eqn:GR_def_ratio}) and (\ref{eqn:GR_tfv_formula}).
Figure \ref{fig:lua_code_example} exemplifies the Lua code checking if an exit event has occurred.  

For further increased performance it is possible to use the LuaJIT implementation \cite{luajit} where 
the Lua code is compiled to machine code during parsing: this allows performance close to what would be obtained by 
defining the states based on compiled code

Finally, the Lua layer can act
as an intermediate proxy between the C++ \gparrep code and 
any other external library, providing the possibility to define states and observables using external software pieces:
one can for example imagine to use tools such as Colvars \cite{colvars_2013} or PLUMED \cite{plumed2_2014}, providing access to an 
extensive ready to use collection of collective variables definitions.\\

\subsection{On the choice of $\tgr$ and $\tcheck$}

As previously mentioned in subsection~\ref{subsec:genParRep} it is not necessary to check at each integration of $t$
if $(X_t \notin \Omega)$ (parallel step) or $(X_t^\refe \notin \Omega)$ (F-V step) as one expects that the exit
time is much larger than $dt$.

And likewise, while it is important to regularly gather the value of the G-R observables $\obs$ in order to obtain 
convergence of Equation~(\ref{eqn:GR_def_ratio}), it is expected that they will not differ that much between time $t$ and $t + dt$:
hence it is interesting to choose $\tgr > dt$.

While the values should be fine tuned for each system, based on our experience we ended up with the following rule of thumb:
one can take $\tgr$ to be $5$ to $100$ 
times the value of $dt$, and $\tcheck$ to be $500$ to $2000$ times $dt$.
This should be adjusted depending on:
(i) 
the amount of calculations involved in the process of defining the state and the G-R observables in the input script, 
and
(ii) the size of the system; for a large solvated protein, if the states and observables only involve distance measures on a few 
atoms, then the time required for performing the dynamics will be comparatively much larger and relatively small values 
of $\tgr \approx 10$ and $\tcheck \approx 250$ can be selected; however, if it involves tracking the length of several dozens of 
hydrogen distances, or counting native contacts, a wise approach would be to choose $\tgr \approx 50$ and $\tcheck 
\approx 1000$.

Furthermore, it should be emphasized that while Equation~(\ref{eqn:ParRep_CorrectedDiscreteTauEscape}) is mathematically valid for any 
values of $\tcheck$
(in the sense that it indeed samples the exit time of the sub-sampled Markov chain $(X_{k \tcheck})_{k\in\mathbb{N}}$),
if $\tcheck$ is taken too large, one may miss an exit event if the process re-enters the same 
state $\Omega$ during the time interval $t \rightarrow t + \tcheck$;
however one may argue that such cases may denote a poor definition of 
the states, and that for states exhibiting strong metastability this should not be an issue. 
\\

%%%%%%%%%%%%%%%%%%%%%%%%%%%%%%%%%%%%%%%%%%%%%%%%%%%%%%%%%%%%%%%%%%%%%%%%%%%%

\section{Results and discussion}
\label{sec:results}

Now that both the algorithm and the software implementation of the Generalized ParRep (in the following denoted as ``\gparrep'') have been 
extensively discussed, let us consider two applications:
the first validates the implementation and consists in a study of the conformational equilibrium of alanine 
dipeptide (subsection~\ref{subsec:ala2}),
while the second investigates the dissociation of the FKBP--DMSO protein--ligand complex
(subsection~\ref{subsec:fkbp}).

In the following, when reporting estimated values of the average exit time $\mathbb{E}(\tau)$ from a metastable state
we will consider the sample average $\bar{\tau}$ over $n$ samples $\{\tau_1,...,\tau_n\}$ 
as
\[ \bar{\tau} = \dfrac{1}{n} \sum_{i=1}^{n} \tau_i .\]
Furthermore the $1-\alpha$ confidence interval for those (close to) exponentially distributed samples is:
\[ \frac{2n\bar{\tau}}{\chi^2_{1-\frac{\alpha}{2},2n}} < \mathbb{E}(\tau) < 
\frac{2n\bar{\tau}}{\chi^2_{\frac{\alpha}{2},2n}} \]
where $\chi^2_{q,\nu}$ is the value of the quantile function of the $\chi^2$ distribution with $\nu$ degrees of freedom 
at level $q$; in the following we chose $\alpha = 0.05$ and therefore report the $95 \%$ confidence interval.

%%%%%%%%%%%%%%%%%%%%%%%%%%%%%%%%%%%%%

\subsection{Conformational equilibrium of the alanine dipeptide}
\label{subsec:ala2}

\newcommand{\tauala}{\tau_{C_{7 \rm eq} \rightarrow C_{7 \rm ax}}}

The blocked alanine dipeptide (Ac-Ala-N-H-Me)
has been used as a validation system for computational studies of
conformational equilibria, and energy landscape reconstruction
and analysis
\cite{ala2_apostolakis_1999,ala2_Chun_2000,ala2_Swope_2004,ala2_Ren_2005,ala2_Jang_2006,ala2_Strodel_2008,ala2_Vedell_2008,ala2_Velez_2009}.
The dipeptide contains several notable structural features, including
the two $(\phi,\psi)$ dihedral angles, NH- and CO-groups capable of
H-bond formation, and a methyl group attached to the $C_\alpha$
atom.
One suitable way to visualize the conformations and the transitions between them is to draw an energy
surface as a Ramachandran plot \cite{Ramachandran_1963}:
when studied \emph{in vacuo} the following two metastable states
are clearly identified:
(i)  $C_{7 \rm eq}$ for $(\phi,\psi) \sim (\ang{-75},\ang{100})$, and 
(ii) $C_{7 \rm ax}$ for $(\phi,\psi) \sim (\ang{60},\ang{-60})$. 

In the following we estimate the mean first passage time $\tauala$ between the two metastable states, using the 
\gparrep algorithm;
accuracy of the method is compared to one long serial Langevin dynamics, as the low 
complexity of this system allows direct numerical simulation of numerous transition events;
finally the influence of some of the \gparrep parameters is also evaluated.\\

%-----------------------------
\subsubsection{MD setup}

The initial configuration of the dipeptide is $(\phi,\psi) = (\ang{-81.0},\ang{70.0})$, i.e. within the most 
populated area of the $C_{7 \rm eq}$ state (see the yellow mark on Figure~\ref{fig:ala2_states_rama} together with a 
representation of the corresponding conformation). The OpenMM system was configured as follows:
the CHARMM 22 all-atoms for proteins and 
lipids force-field including CMAP corrections \cite{charmm22ff_1998,charmm22ff_2004} was used;
dynamics was performed using a Langevin integrator (time-step of $dt = 2 \rm~fs$, friction of $\gamma = 2 
\rm~ps^{-1}$), thermostated at a temperature of $T = 300 \rm~K$;
the non-bonded interactions were evaluated using a non-periodic cutoff scheme up to a distance of $1.6 \rm~nm$;
and bonds involving hydrogens are constrained to a value of $\pm 10^{-3}~\%$ of their original distance.\\

%-----------------------------
\subsubsection{\gparrep setup}

\begin{figure}[h!]
\centering
\includegraphics[width=0.85\linewidth]{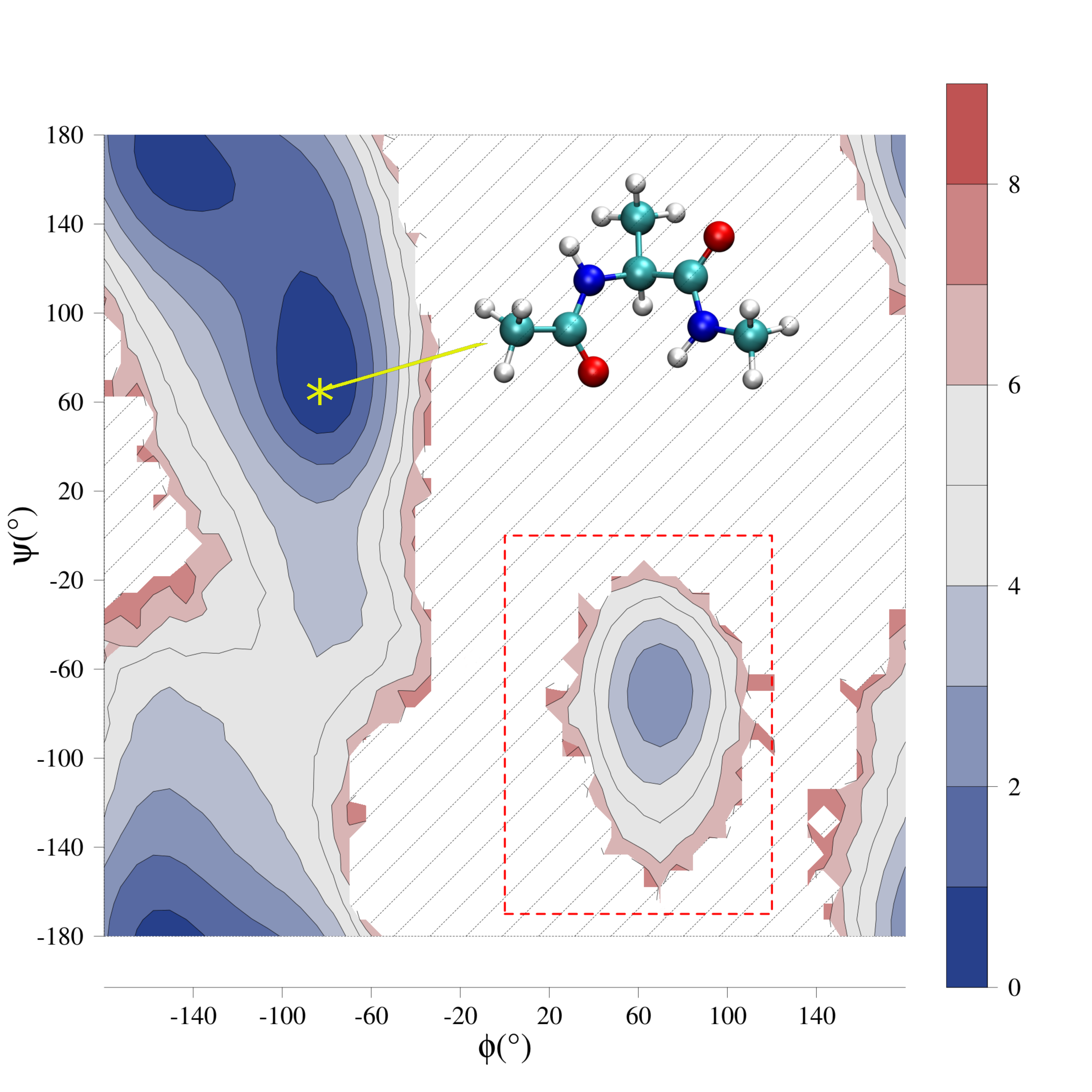}
\caption{Alanine dipeptide : definition of ParRep domains based on a free energy surface 
(color coded, in kcal/mol, dashed oblique lines correspond to unsampled areas), constructed from the long MD 
reference simulation. The red rectangle corresponding to $\phi \in [0;120]~\text{and}~\psi \in [-170;0]$ 
is used as a threshold defining the $C_{7 \rm ax}$ state (see subsection~\ref{subsec:ala2} for details).
The yellow cross corresponds to the starting configuration for either \gparrep or serial MD simulations.
\label{fig:ala2_states_rama}
}
\end{figure}

The procedure for defining the states may have to be adapted for each force-field and in the following we assume the 
use of the aforementioned CHARMM22 force-field.
Figure~\ref{fig:ala2_states_rama} is a Ramachandran plot based free energy surface built from preliminary serial 
Langevin MD simulation: it illustrates how the ParRep states were a priori defined.

One can see that in the upper left quarter of the plot two close stable conformations indeed coexist, separated by a
low energetic barrier of $1$ to $2$ kcal/mol, which is comparable to the product $k_BT$: hence it was decided to 
combine those two minima together, as they do not constitute alone a valid metastable target for applying the 
ParRep method (Refs. \cite{ala2_Chun_2000,ala2_Strodel_2008,ala2_Velez_2009} indeed suggest that
the transition between those two wells is of $2.7$, $3.0$ and $4.05~\rm ps$, respectively).
Therefore the conformational equilibrium
of the dipeptide is modeled using a two states definition:

\begin{enumerate}
\item The $C_{7 \rm ax}$ ParRep state corresponds to the well for which $\phi > \ang{0}$ and $\psi < \ang{0}$, i.e. 
the lower right quarter of Figure~\ref{fig:ala2_states_rama}; it was decided to model this state using the following
rectangular domain:
\[ \phi \in [0;120]~\text{and}~\psi \in [-170;0]  \]
represented as a red rectangle in Figure~\ref{fig:ala2_states_rama}.

\item 
The $C_{7 \rm eq}$ ParRep state consists in the set of all configurations not falling within the red rectangle: it is therefore the 
complement of the state $C_{7 \rm ax}$.
\end{enumerate}

Therefore this setup corresponds to a two states partition of the configuration space projected onto a Ramachandran plot.

Concerning the Fleming-Viot procedure, four Gelman-Rubin observables $\obs$ are considered for tracking the convergence 
to the QSD: the total potential energy $V(q)$, the kinetic energy $K(p) = \frac{1}{2} p^T M^{-1} p$, and the value of 
the $\phi$ and $\psi$ dihedral angles acting here as collective variables.
The tolerance criterion $\tol$ is fixed per simulation to a given value which is the same for 
each observable (the influence of $\tol$ is investigated below for a range of values).
The value of $\tgr$ (accumulation of observables) was set to $10 \times dt$ (i.e. $20~\rm fs$) as the
observables are not computationally expansive to calculate, and the test  $(X_t^\refe \notin \Omega)$
is performed at $\tcheck = 250 \times dt$ (i.e. $0.5~\rm ps$) during the Convergence step, but at $\tcheck = 2500 \times dt$ during the 
Parallel 
dynamics step, which corresponds to $5~\rm ps$, in order to maximize the CPU time spent in the Langevin dynamics.\\

%-----------------------------
\subsubsection{Discussion}

In the following the distribution of the \gparrep sampled values $\tauala$ are compared to results obtained when performing
a long reference dynamics (denoted as \emph{reference MD} in the following), consisting in a Langevin dynamics simulation of
a total length of $162~\mathrm{\mu s}$.
As a result, $533$ $\tau_{C_{7 \rm eq} \rightarrow C_{7 \rm ax}}$ events were sampled, and 
$\mathbb{E}(\tau_{\rm MD ref}) = 304.47~\rm ns$ is obtained, while the confidence interval
is $280.19~\text{ns} < \mathbb{E}(\tau_{\rm MD ref}) < 332.07~\text{ns}$ (see Table~\ref{tab:ala2_stats} for a summary);
in Ref. \cite{ala2_Strodel_2008} the authors estimate $\tauala$ to be of $353 \rm~ns$ (no provided error estimate), in agreement with
this value.

First, the possibility to obtain an accurate estimate of $\tauala$ by generating a relatively small number $n$
of samples is investigated:
the number of \gparrep replicas was set to $N = 224$, and the number $n$ of samples of $\tauala$ generated was $\{31,39,40\}$
for respective tolerance levels of $\tol = \{0.01,0.025,0.05\}$ (red, green and blue solid lines; less samples were collected for $\tol = 
0.01$ because of the higher computational effort required for lower tolerance values).

\begin{figure}[h!]
\centering
\includegraphics[width=0.85\linewidth]{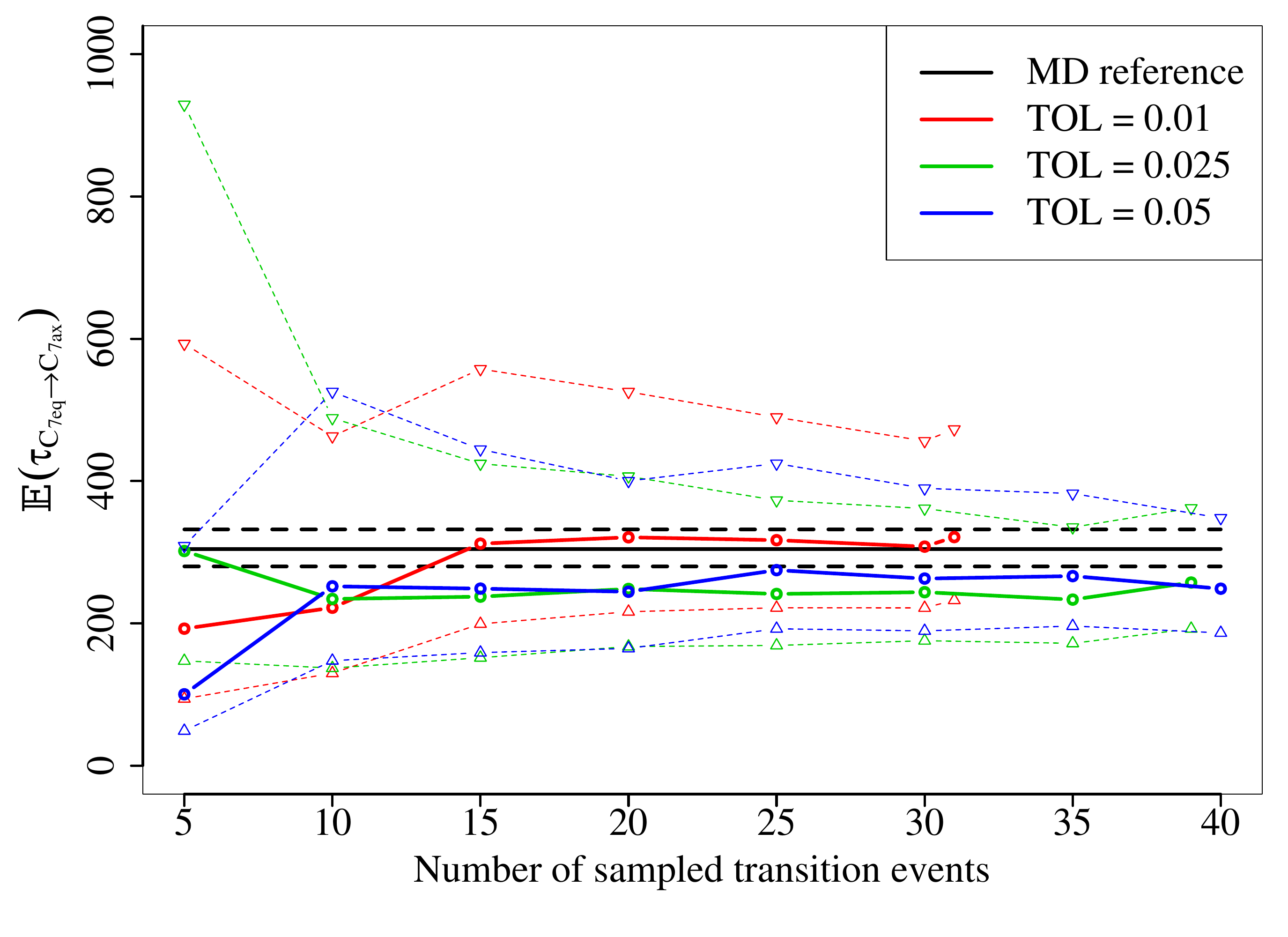}
\caption{
Convergence of $\mathbb{E}(\tauala)$ (plain lines) to the MD reference (black lines), for \gparrep sampled values (red, green and blue 
lines), 
when 
considering only the first $m$ of the $n = \{31,39,40\}$ samples (abscissa), for $\tol$ levels of respectively $\{0.01,0.025,0.05\}$.
Dashed lines correspond to the $95 \%$ confidence interval; The number of \gparrep replicas was fixed at $N = 224$.
\label{fig:ala2_convergence_fewpoints}
}
\end{figure}

The convergence to the MD reference (black lines) can be visualized on Figure~\ref{fig:ala2_convergence_fewpoints} where 
$\mathbb{E}(\tauala)$ (solid lines) and the corresponding confidence interval (dashed lines) are given for the three ParRep
simulations, when considering only the subset of the first $m$ sampled values;
the red line ($\tol = 0.01$) quickly converges to the same distribution than 
the MD reference, while higher values of $\tol$ appear to converge to a different distribution underestimating
$\mathbb{E}(\tau)$ (see also numerical values in Table~\ref{tab:ala2_stats}): this suggests that convergence to the QSD is only 
obtained for a value of $\tol = 0.01$ when studying the $C_{7 \rm eq} \rightarrow C_{7 \rm ax}$ transition.

\begin{figure}[h!]
\centering
\includegraphics[width=0.85\linewidth]{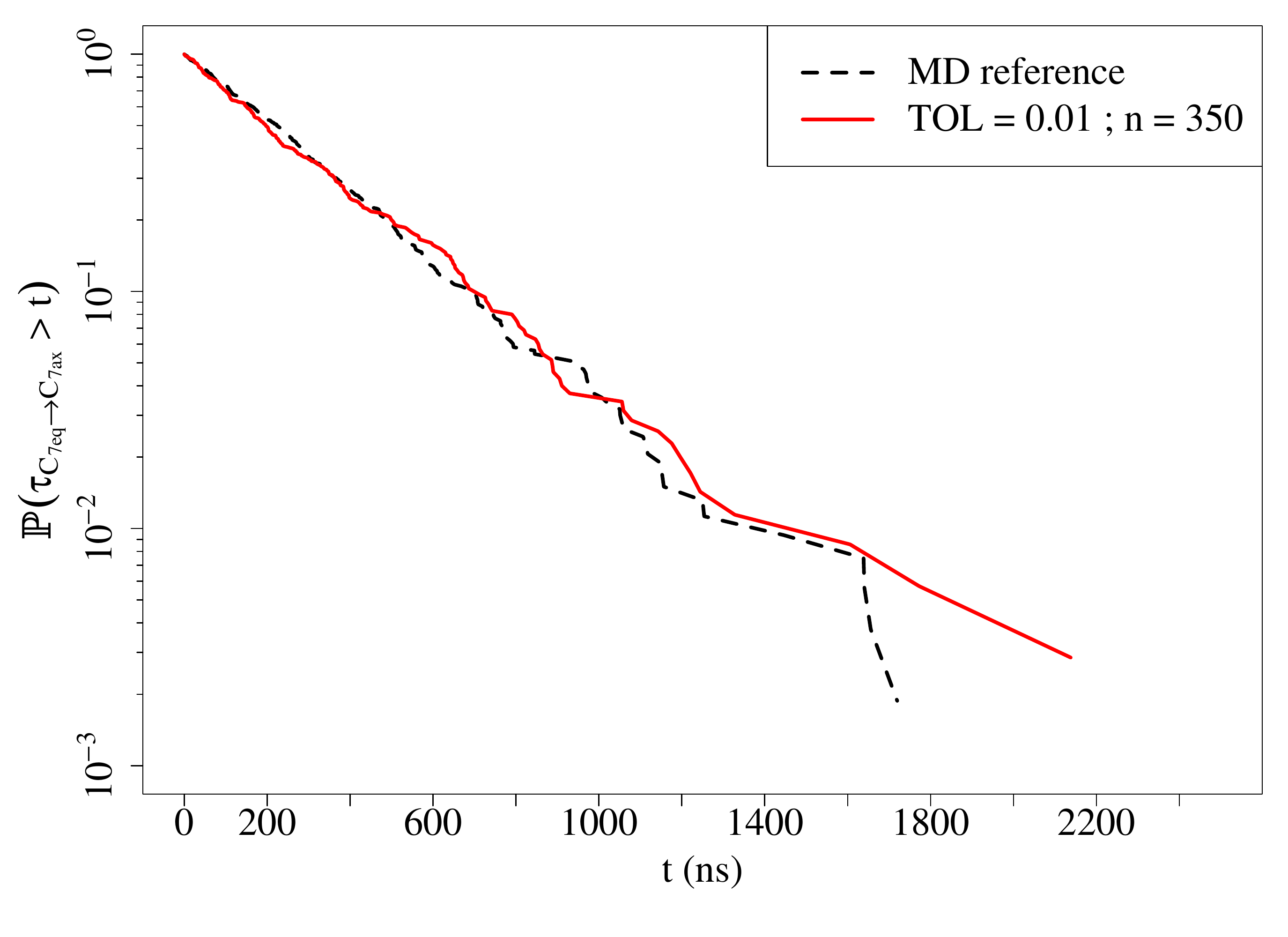}
\caption{
Distribution of \gparrep sampled values of $\tauala$ for $\tol = 0.01$,
for $N = 224$, and a number of $\tau$ values generated $n = 350$.
\label{fig:ala2_distrib_manypoints}
}
\end{figure}

\begin{figure}[h!]
\centering
\includegraphics[width=0.85\linewidth]{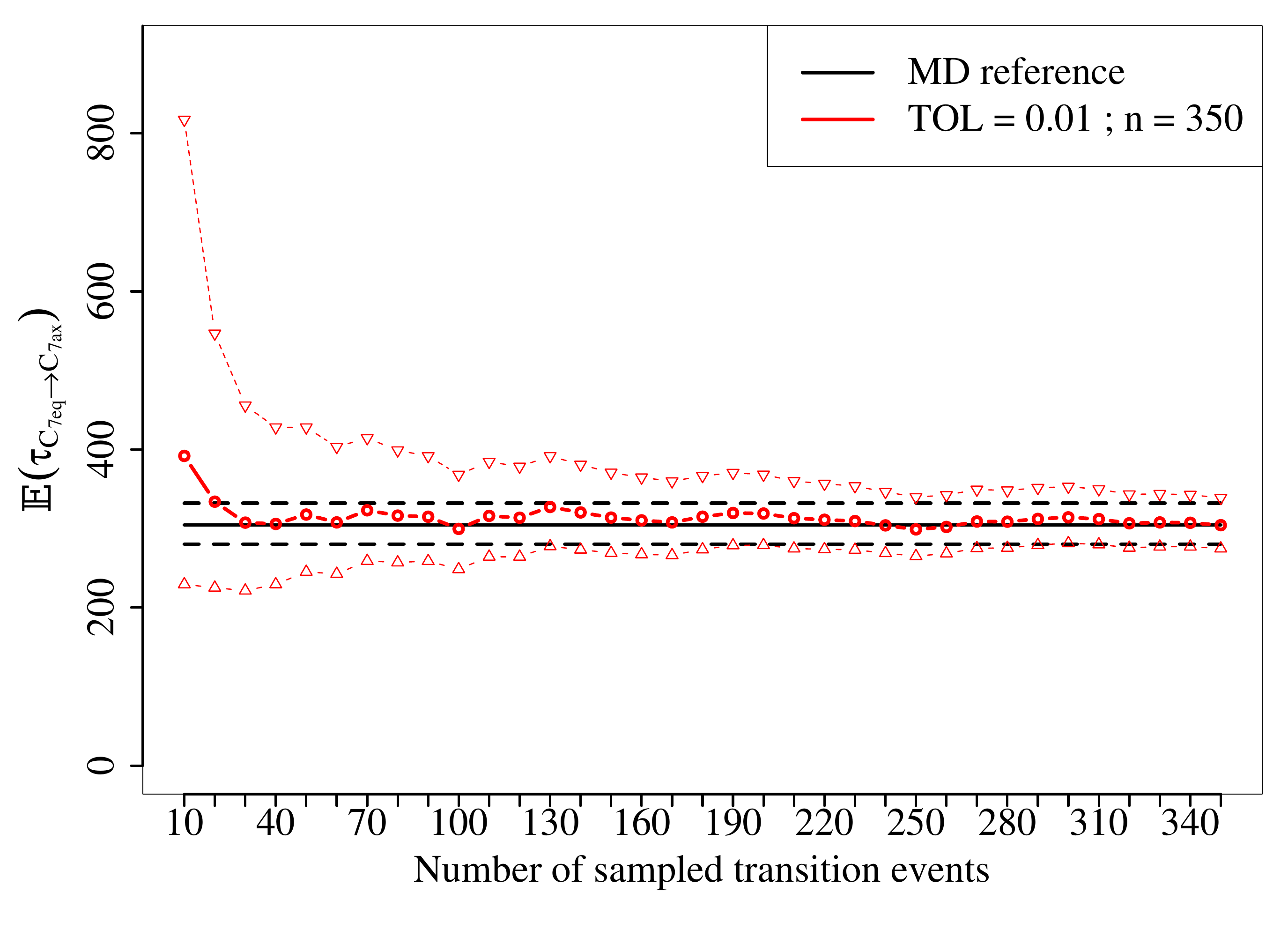}
\caption{
Convergence of $\mathbb{E}(\tauala)$ (solid lines) for \gparrep sampled values from Figure~\ref{fig:ala2_distrib_manypoints}, when 
considering only the first $m$ sampled values among $n = 350$ (abscissa). Dashed lines correspond to the $95 \%$ confidence interval.
\label{fig:ala2_convergence_manypoints}
}
\end{figure}

Now that a value of $\tol = 0.01$ appears to be accurate enough, one can collect more samples $n$ in order
to verify that the distribution of the exit times converges to the one obtained with the reference MD simulation.
In Figure~\ref{fig:ala2_distrib_manypoints}, the distribution of $n = 350$ samples generated for a level of $\tol = 0.01$ and using
$N = 224$ is illustrated: this was done by building an empirical \emph{complementary cumulative distribution function}
(in the following referred to as $ccdf$) using the $n$ samples, and 
it provides an estimate of the probability that $\tauala$ is higher than a given value $t$ (i.e. $\mathbb{P}(\tauala > t)$), with by 
definition $\mathbb{P}(\tauala > 0) = 1$).
One can see from Figure~\ref{fig:ala2_distrib_manypoints} that the \gparrep
and MD distributions are in really good agreement for $t \in [0;1500]~\rm ns$, where a quasi linear 
function $t \mapsto \ln \mathbb{P}(\tauala > t)$(i.e. exponential law) is observed
(we ignore the area for $t > 1500~\rm ns$, i.e. the low 
probability tail of
the distribution corresponding to large values of $\tauala$, where the MD simulation lacks samples for performing a meaningful 
comparison).
This observation is confirmed by looking at Figure~\ref{fig:ala2_convergence_manypoints}: the convergence of \gparrep
samples to the MD reference is observed, both for the mean value and the confidence interval, for increasing values of 
$n$.
\\  

%-----------------------------
\subsubsection{Distribution of the F-V estimated value $\tfv$}

\begin{figure}[h!]
\centering
\includegraphics[width=0.85\linewidth]{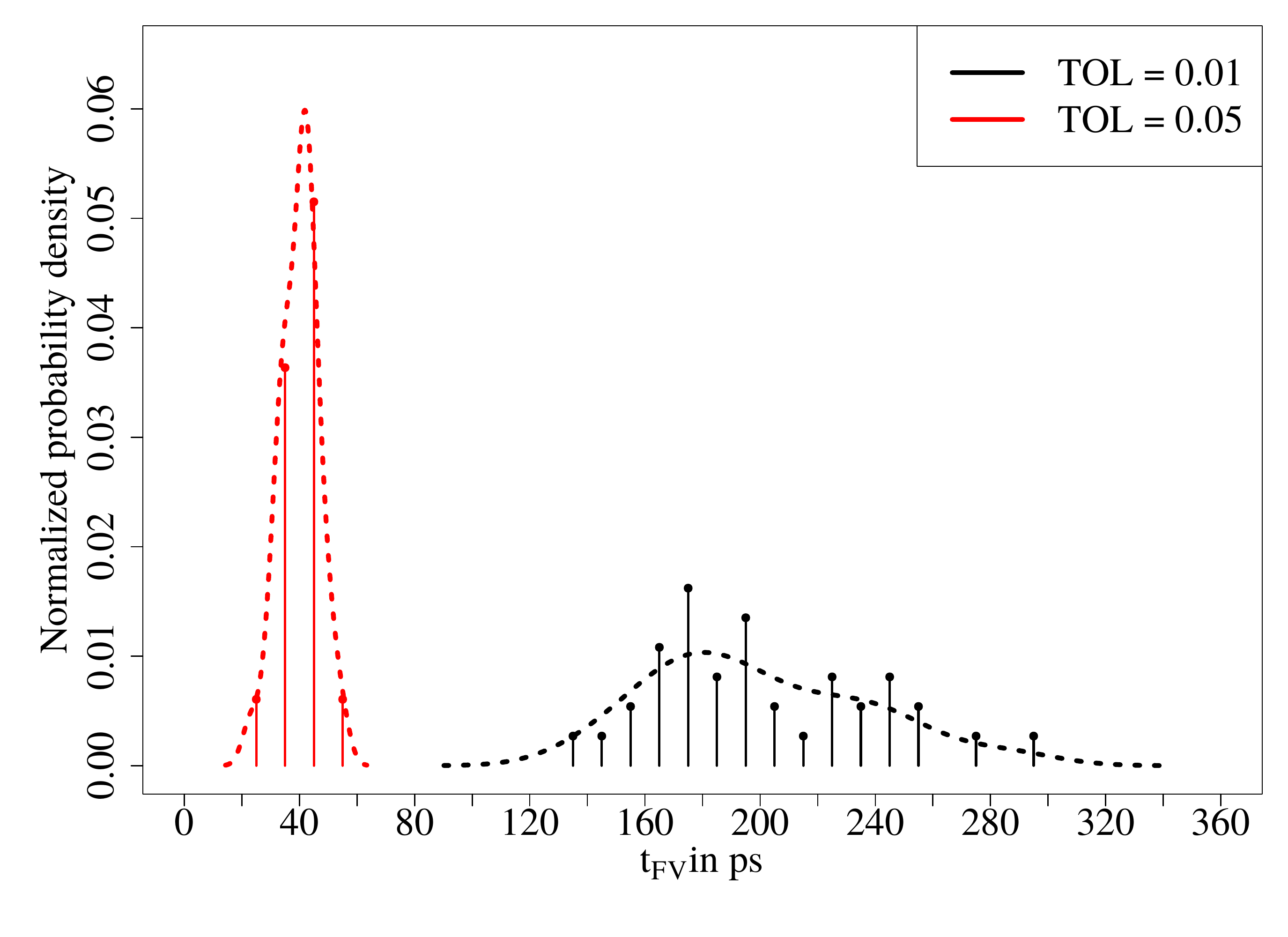}
\caption{
Histogram distribution (vertical lines) of $\tfv$, obtained
for two different tolerance levels of $\tol = \{0.01,0.05\}$ (two datasets from Figure~\ref{fig:ala2_convergence_fewpoints}): $\tfv$ 
corresponds to the simulation time before one assumes that the samples are distributed 
according to the QSD (see subsection~\ref{subsec:genParRep}).
The dashed lines correspond to a Kernel Density Estimation \cite{kde_rosenblatt_1956,kde_parzen_1962} smoothing.
\label{fig:ala2_tfv_distrib}
}
\end{figure}

One last interesting quantity to collect is the time $\tfv$ necessary for the convergence of the observables $\obs$
defined by Equation~(\ref{eqn:GR_tfv_formula}) (see subsection~\ref{subsec:genParRep} for more details).
Figure~\ref{fig:ala2_tfv_distrib} provides
the histogram distribution of $\tfv$ for 
two of the datasets from Figure~\ref{fig:ala2_convergence_fewpoints}, together with a Kernel Density Estimate 
smoothing \cite{kde_rosenblatt_1956,kde_parzen_1962} (dashed lines). 
For a tolerance of $0.05$ one observes that $\tfv \approx 40 \rm~ps$ and appears to follow a normal distribution; for $\tol = 0.01$
it seems that the distribution is bimodal, with a major mode at $\tfv \approx 180 \rm~ps$ and a minor mode at $\tfv \approx 240 \rm~ps$.
While it is difficult to argue how the distribution of $\tfv$ ideally looks like, one should remember that the 
$C_{7 \rm eq}$ is defined as the large funnel on the left ($\phi < \ang{0}$) side of Figure~\ref{fig:ala2_states_rama}, and
that therefore it encompasses the whole range of the possible $\psi$ values, meaning that $\psi$ will be the slowest observable to converge;
it is thus expected that the value of $\tfv$ will be large for conservative tolerance levels ($\tol \to 0$), and that depending on how the 
F-V workers randomly diffused on the $(\phi,\psi)$ surface, its distribution will be broad, possibly multimodal. Hence the dispersion for 
$\tol = 0.01$ in Figure~\ref{fig:ala2_tfv_distrib} appears coherent, while the homogeneous distribution for $\tol = 0.05$ probably 
indicates that the F-V workers did not diffuse far enough from their starting point in $C_{7 \rm eq}$: they are therefore still distant 
from what the QSD would be, and this explains why $\mathbb{E}(\tauala)$ never converged to the result obtained by direct numerical 
simulation when $\tol = 0.05$.

Figure~\ref{fig:ala2_tfv_distrib} emphasizes one of the main advantages of the \gparrep algorithm versus the original method, i.e. the fact 
that $\tfv$ is calculated on the fly, and thus adjusted to the initial condition within the state, whereas the original 
algorithm required a fixed user defined value 
after which it was assumed that the 
QSD was reached: indeed, one can see that for $\tol = 0.01$ (which seems necessary to be sufficiently close to the exact QSD), the 
distribution of $\tfv$ is 
spread over an interval going from $120$ to $300~\rm ps$;
it is therefore obvious that choosing a priori a decorrelation time of $120 \rm~ps$ would result in a bias as this value appears to be far 
below the time it takes to reach the QSD for some initial conditions; and on the contrary choosing a decorrelation time of $300 \rm~ps$ 
would ensure 
quasi-convergence to the QSD for most of the initial conditions, but at the cost of an unnecessary long (and then costly) decorrelation 
step for some of the initial conditions.\\

\begin{table}[h!]
\centering
\caption{
Summary of the estimated value of $\mathbb{E}(\tauala)$ and of the corresponding $95 \%$
confidence interval
for data presented in Figures \ref{fig:ala2_convergence_fewpoints} to \ref{fig:ala2_convergence_manypoints}.
The \gparrep results ($N = 224$) appear to converge accurately for a value of $\tol = 0.01$. 
\label{tab:ala2_stats}
}
\begin{tabular}{c|c|c|c|c|c}
	 Method  &  $n$  &  $N$  & $\tol$  & $\mathbb{E}(\tau)$ (ns)    &  Confidence interval (ns) \\ \hline
	MD ref.  & $533$ &  ---  &   ---   &     $\mathbf{304.47}$      &  $(280.19,332.07)$   \\
	\gparrep & $40$  & $224$ & $0.05$  &          $248.72$          &  $(186.60,348.14)$   \\
	\gparrep & $39$  & $224$ & $0.025$ &          $257.37$          &  $(192.45,361.94)$   \\
	\gparrep & $31$  & $224$ & $0.01$  &          $321.26$          &  $(232.54,472.83)$   \\
	\gparrep & $350$ & $224$ & $0.01$  &     $\mathbf{304.22}$      &  $(274.70,338.78)$   \\
\end{tabular} 
\end{table}

%-----------------------------
\subsubsection{Performance}

The last point to discuss concerns the performance of the \gparrep method and particularly the speedup compared with the reference 
serial Langevin dynamics.

In Table~\ref{tab:ala2_benchmark} benchmarking data is reported for the simulations from Figure~\ref{fig:ala2_convergence_fewpoints}; the 
fifth column reports the calculated effective speedup which is compared to the maximum 
possible speedup $N = 224$; the sixth column reports the ratio between the effective speedup (see Table's caption for 
methodology) and the maximum possible speedup (hence a value of $100 \%$ would indicate a perfect linear speedup).

One can see that for a large tolerance of $0.05$ a value of $189$ is obtained, i.e. $84~\%$ of the maximum possible value; and 
for a more conservative tolerance criterion of $0.01$ this falls to $156$ i.e. $70~\%$ of $N = 224$: this illustrates the cost of an 
accurate convergence step which, as seen in the previous section, is the key for obtaining accurate results.

Considering the reduced size of the system ($22$ atoms) and the fact that during the F-V procedure the MD engine code is interrupted every 
$10 \times dt$ for collecting the value of the G-R observables, this speedup is an impressive result;
although slightly higher values may be obtained by tunning further the values of 
$\tgr$ and $\tcheck$, there is probably little space for optimization for such a small test-case system: therefore a more detailed 
performance analysis will be performed in the next subsection for the protein--ligand system.\\

\begin{table}[h!]
\centering
\caption{
Benchmarking data for the three datasets from Figure~\ref{fig:ala2_convergence_fewpoints} ($N = 224$,
$n = \{31,39,40\}$ and $\tol = \{0.01,0.025,0.05\}$). Each replica runs on $P = 1$ CPU cores.
The wall-clock time (column~$2$) is taken as the time elapsed from the beginning to end of the execution of the program, it includes 
both computations and communications time.
The speed (ns/day, column~$4$) is obtained by dividing the total simulation clock $\tsim$ (column~$3$) by the value of the wall-clock time 
(in days).
The effective speedup (column~$5$) corresponds to column~$4$ divided by the performance of a serial MD reference (evaluated as $921$ 
ns/day by independent tests on the same architecture).
The ratio between the effective and maximum possible speedup (by definition equal to $N = 224$) is given as a percentage in column~$6$:
a theoretical value of $100\%$ would correspond to the maximum possible speedup.
\label{tab:ala2_benchmark}
}
\begin{tabular}{l|c|c|c|c|c}
	$\tol$  & WT(s)  & $\tsim$(ns) & Speed(ns/day) & Eff. speedup & (Eff./Max.) \\ \hline
	$0.01$  & $6015$ &   $10008$   &   $143752$    &    $156$     &   $70 \%$   \\
	$0.025$ & $5239$ &   $10103$   &   $166609$    &    $181$     &   $80 \%$   \\
	$0.05$  & $4973$ &   $10032$   &   $174296$    &    $189$     &   $84 \%$
\end{tabular}
\end{table}

%%%%%%%%%%%%%%%%%%%%%%%%%%%%%%%%%%%%%
\subsection{Dissociation of the FKBP--DMSO protein--ligand system}
\label{subsec:fkbp}

\newcommand{\tauoff}{\tau_{\rm off}}

\begin{figure}[h!]
\centering
\subfigure[~Bound state]  {\centering\includegraphics[width=0.45\linewidth]{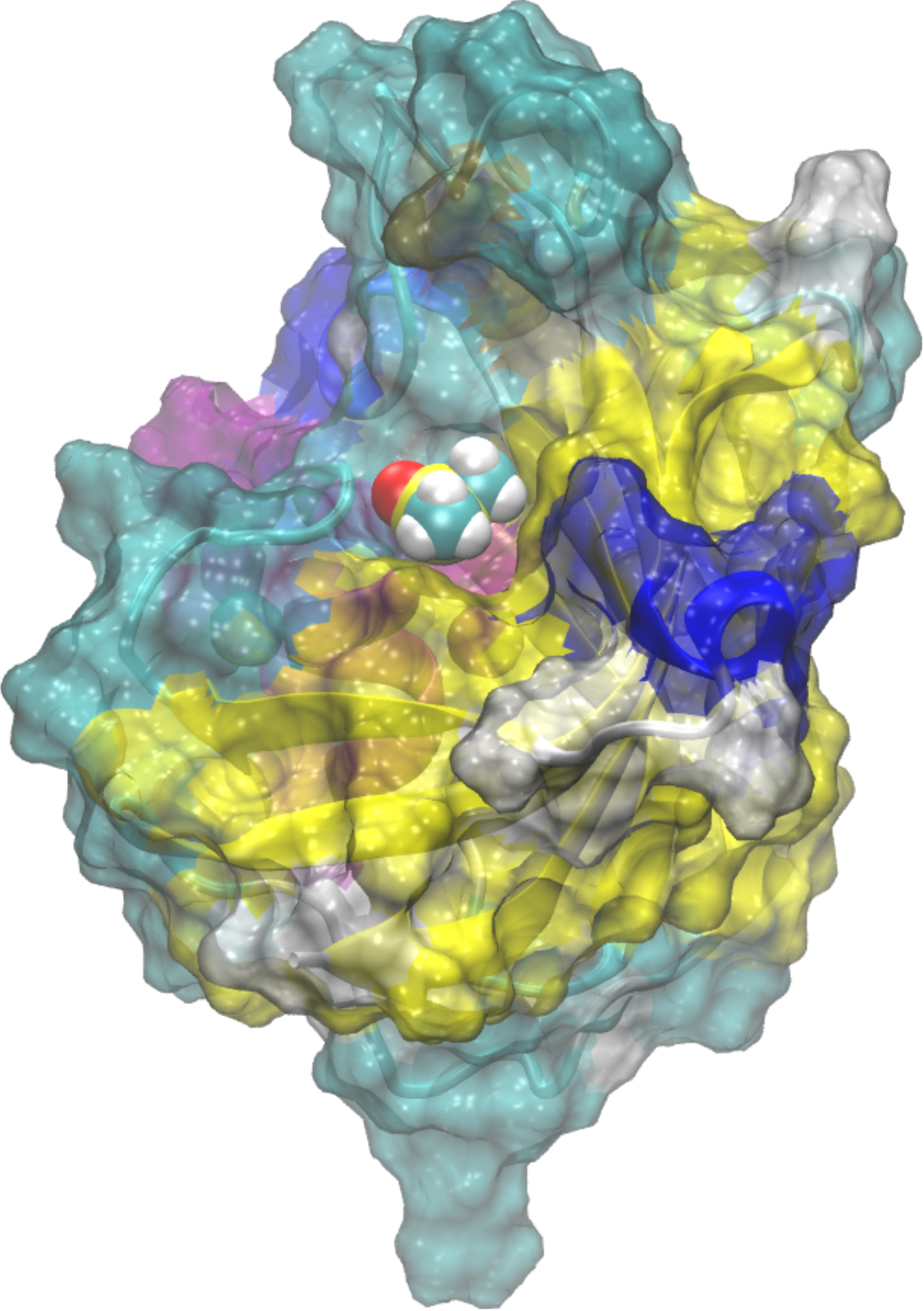}}
\subfigure[~Unbound state]{\centering\includegraphics[width=0.45\linewidth]{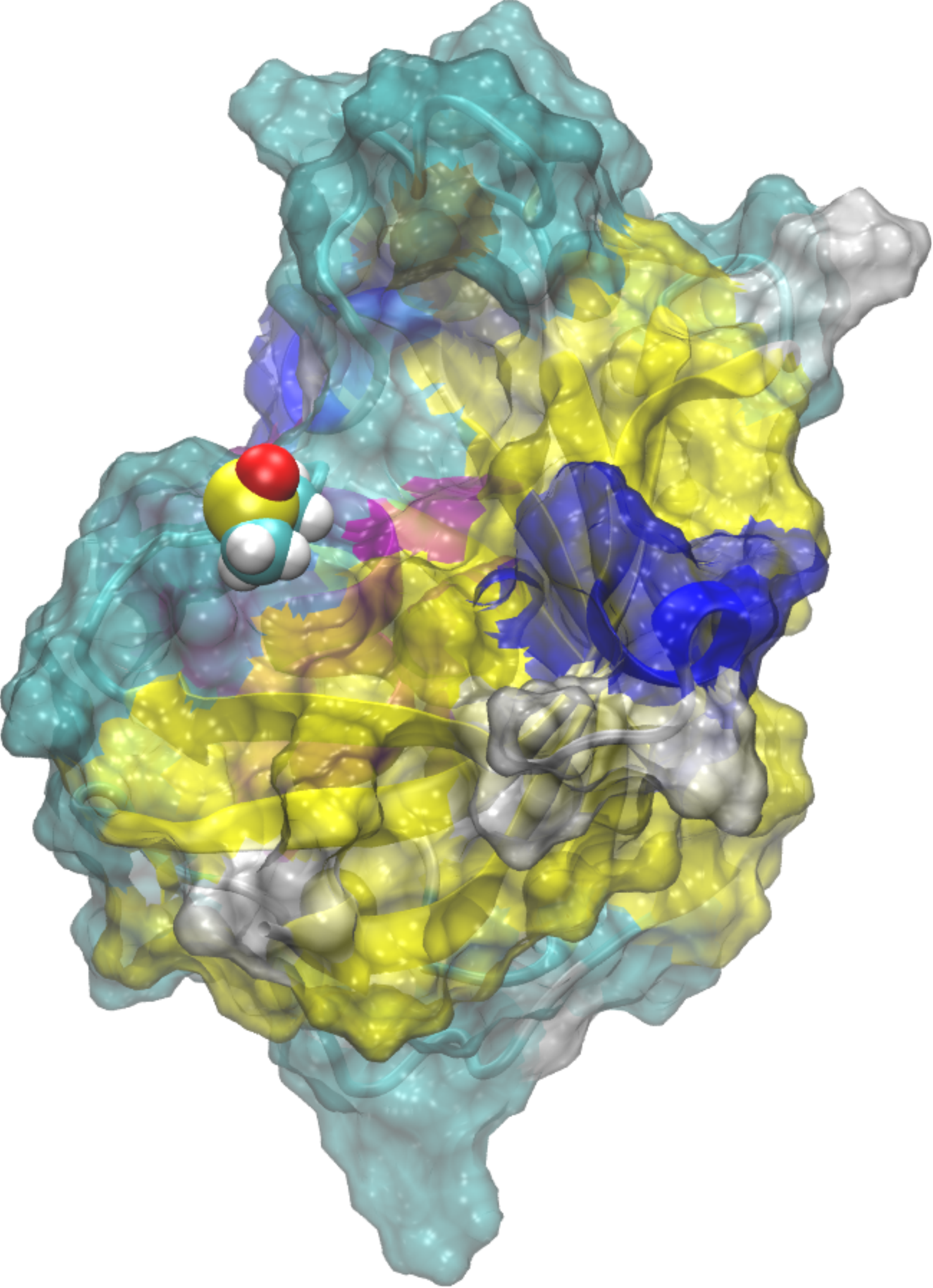}}
\caption{Illustration of the FKBP--DMSO complex, corresponding to the RCSB-PDB entry ``1D7H" : on the left the 
undissociated (``bound") state used as starting configuration for all the simulations; on the right 
the target dissociated (``unbound") state characterized by a $\tauoff$ dissociation time.
\label{fig:FKBP_b_u}
}
\end{figure}

After validation of the \gparrep algorithm on the alanine dipeptide, we would like to demonstrate its efficiency on a 
larger protein--ligand system:
the aim is to sample the dissociation time $\tauoff$ between the \emph{bound} and \emph{unbound} states
of the FKBP--DMSO complex (see Figure~\ref{fig:FKBP_b_u}).
The FKBP protein (also known as the FK506 binding protein) have a role in the folding of other proteins containing proline 
residues \cite{Siekierka_fkbp_1989}; in the human body the FKBP12 protein binds to the tacrolimus molecule (and derivatives), an 
immunosuppressant drug used to reduce organ rejection after an organ transplant \cite{Wang_fkbp_1994}.
Because of this important role, both experimental studies \cite{burkhard_fkbp_2000} and molecular dynamics 
simulations \cite{huang_fkbp_free_2011,huang_fkbp_small_2011,xu_fkbp_2016} were performed for evaluating the affinity of the FKBP protein 
to multiple ligands; these include the DMSO (Dimethyl-sulfoxide), a small molecule with anti-inflammatory, antioxidant and analgesic 
activities \cite{Smith_fkbp_1998}, and often used in topical treatments because of its membrane-penetrating ability, which enhances the 
diffusion of other substances through the skin \cite{Malik_fkbp_1995}.

%-----------------------------
\subsubsection{MD setup}
The initial configuration was taken from the RCSB-PDB entry ``1D7H";
the AmberTools17 \cite{amber_2017} software suite was used for setting up an implicit solvent
input configuration (using the OBC \cite{obc2_2004} model II): first, parameters for the DMSO ligand were retrieved
from the GAFF \cite{gaff_2004} force-field using the antechamber program; then parameters for the protein are taken from
the ff14SB \cite{ff14sb_2015} force-field;
dynamics was performed using a Langevin integrator (time-step of $dt = 2 \rm~fs$, friction of $\gamma = 2 
\rm~ps^{-1}$), thermostated at a temperature of $T = 310 \rm~K$;
the non-bonded interactions were evaluated using a non-periodic cutoff scheme up to a distance of $1.6 \rm~nm$;
and bonds involving hydrogens are constrained to a value of $\pm 10^{-3}~\%$ of their original distance.

Before running ParRep simulations, the system was equilibrated for $1.0$ ns, with the DMSO's center of mass being position-constrained 
within $0.36$ nm of its original crystallographic position (force constant of 50 kJ/mol/nm$^2$).
\\

%-----------------------------
\subsubsection{\gparrep setup}

\begin{figure}[h!]
\centering
\subfigure[~DMSO in its binding cavity]{\centering\includegraphics[width=0.45\linewidth]{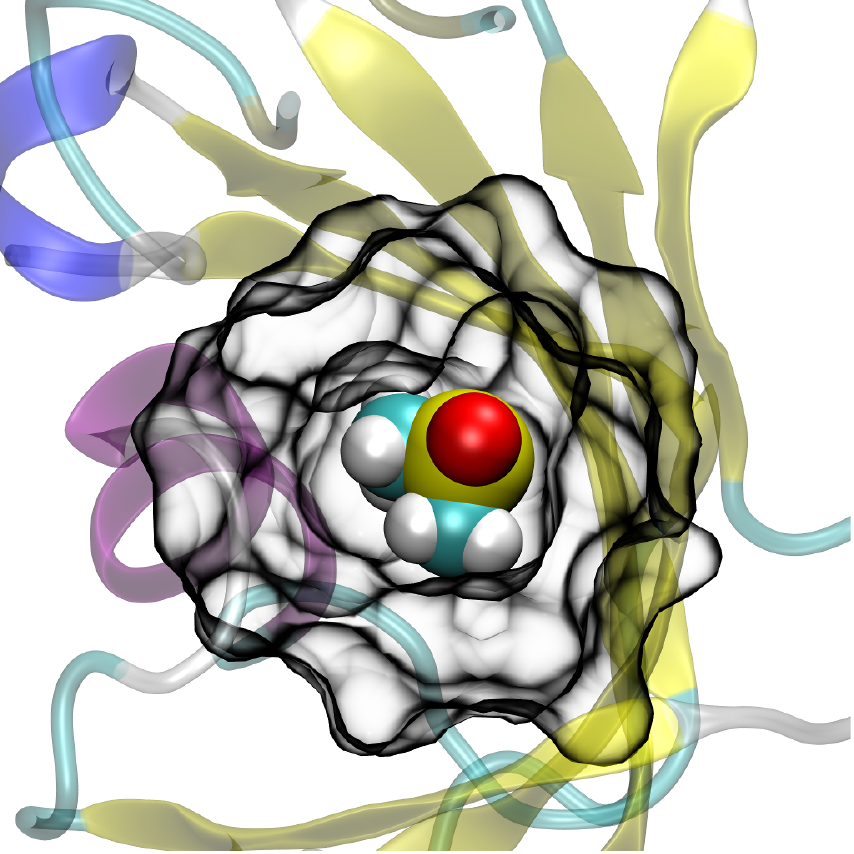}}
\subfigure[~Tracked distances]         {\centering\includegraphics[width=0.45\linewidth]{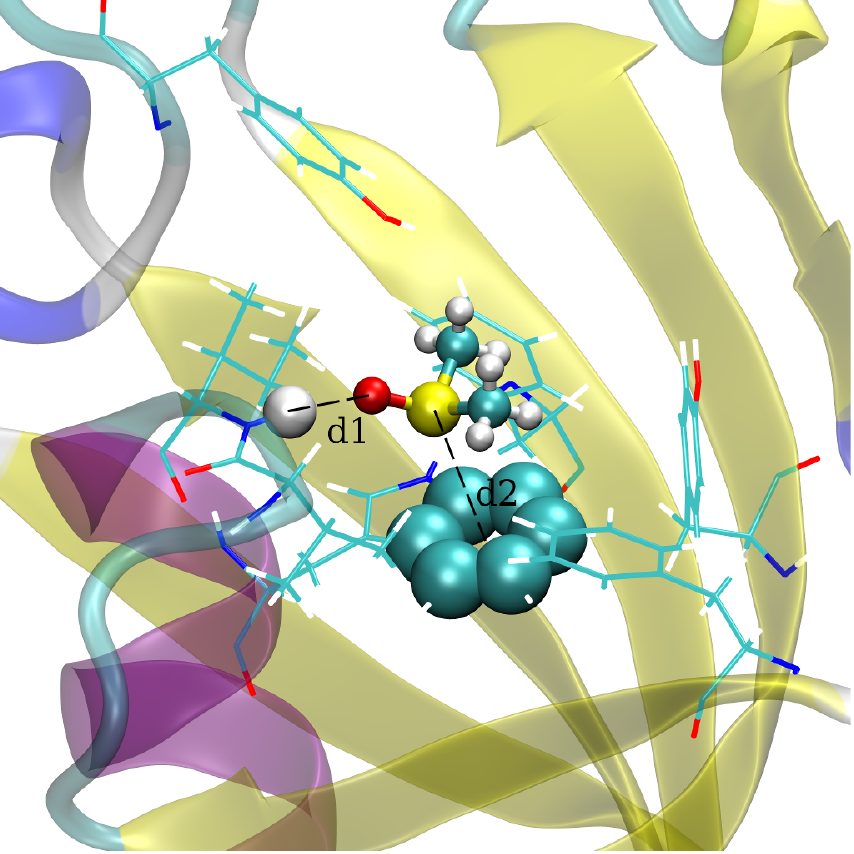}}
\caption{On the left a closer view of the DMSO ligand in its binding cavity: favorable interactions between the O or S 
atoms and surface residues, together with the little available space around the ligand, are responsible for 
metastability.
On the right, surrounding residues within the cavity are represented: in order to detect the dissociation event two 
distances are tracked for defining the bound state (see the \gparrep setup paragraph for details).
\label{fig:FKBP_cavity_dists}
}
\end{figure}

For defining the ParRep states, we used the following procedure, inspired from Refs. \cite{huang_fkbp_free_2011,xu_fkbp_2016}:
a closer view at the ligand binding cavity (see Figure~\ref{fig:FKBP_cavity_dists} (a)) reveals a dense packing with only little
available space around the ligand, and one expects that the sulfur and oxygen atoms will interact favorably via non-bonded interactions
with the surface of the protein; when observing in details the residues surrounding the DMSO (see Figure~\ref{fig:FKBP_cavity_dists} (b) 
corresponding to the RCSB-PDB structure obtained from X-ray 
diffraction \cite{burkhard_fkbp_2000}), one can see
favorable interaction of the O atom with residue ILE-$56$ and of the S atom with residue TRP-$59$.

Hence we used for defining the \gparrep metastable ``bound state'' (denoted by $b$) a criteria based on the distances $d_1$ and $d_2$ as 
illustrated in Figure~\ref{fig:FKBP_cavity_dists} (b):
$d_1$ corresponds to the distance between ligand's oxygen and the hydrogen amide of residue
ILE-$56$;
and $d_2$ corresponds to the distance between ligand's sulfur and the center of mass
of the carbons forming the ring of residue TRP-$59$.
The DMSO is considered to be in the $b$ state when any of $d_1$ or $d_2$ is less than $1.2$~nm, and the 
``unbound state'' (denoted by $u$) is simply defined as configurations where both distances are larger than $1.2$~nm.

\begin{figure}[h!]
\centering
\includegraphics[width=0.85\linewidth]{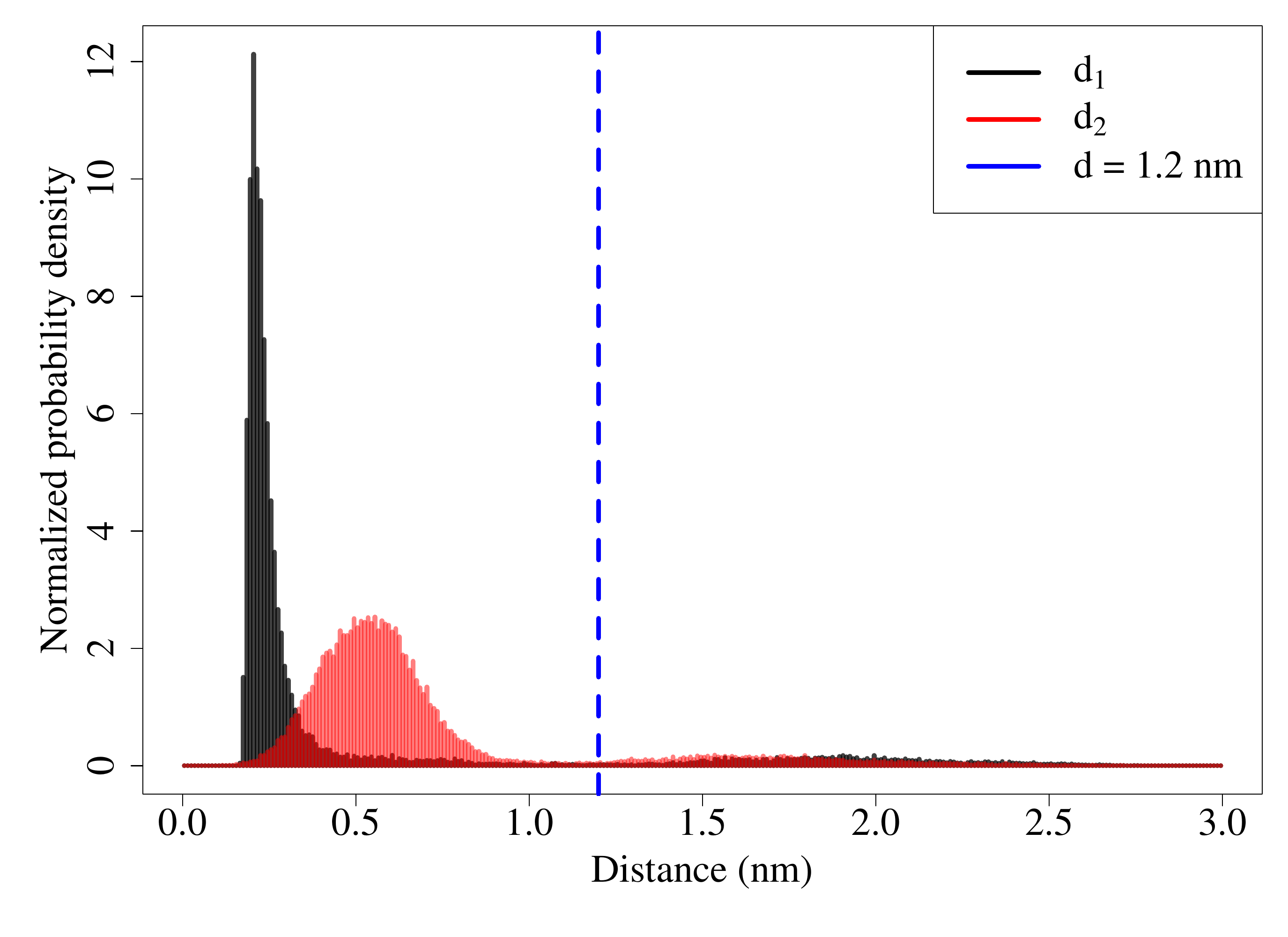}
\caption{
Histogram distribution of the  two d$_1$ and d$_2$ distances as defined in Figure~\ref{fig:FKBP_cavity_dists} (b), for $30~\rm ns$ of plain 
Langevin dynamics.
\label{fig:FKBP_dhist}
}
\end{figure}

One may wonder whether this distance threshold $d < 1.2~\rm nm$ has a physical meaning:
Figure~\ref{fig:FKBP_dhist} shows a histogram distribution of the two distances $d_1$ and $d_2$, for a $30~\rm ns$ Langevin dynamics,
it appears that the threshold of $1.2$~nm corresponds to rarely sampled configurations, far enough
from the top of the two distributions ($\approx 0.25$ and $\approx 0.55$~nm, respectively),
but still closer than the distance range around
$d \ge 1.4$~nm, corresponding to unbound states.
This threshold therefore appears to approximately correspond to a boundary between the 
$b$ and $u$ states.

Concerning the Fleming-Viot procedure, once again $4$ observables were selected in order to track convergence to the QSD:
the two first are the aforementioned distances $d_1$ and $d_2$; the third one is the distance between 
the center of mass of the DMSO and the center of mass of the protein; the last one is the root mean square velocity of the DMSO ligand. 
The levels of $\tol$ were set to the same value for each of the observable, and this value
will be the main \gparrep parameter discussed below.
The value of $\tgr$ was set to $50 \times dt$, and the test  $(X_t \notin \Omega)$
is performed with a period $\tcheck = 1000 \times dt$, both during the Convergence step and the Parallel dynamics step.\\

%-----------------------------
\subsubsection{Discussion}

In the following we will compare the \gparrep results to Ref. \cite{xu_fkbp_2016},
where the authors performed long explicit water MD simulations using the CHARMM 27 force-field and where a value of $\mathbb{E}(\tauoff) = 
2.2~\rm ns$ is reported.

\begin{figure}[h!]
\centering
\includegraphics[width=0.85\linewidth]{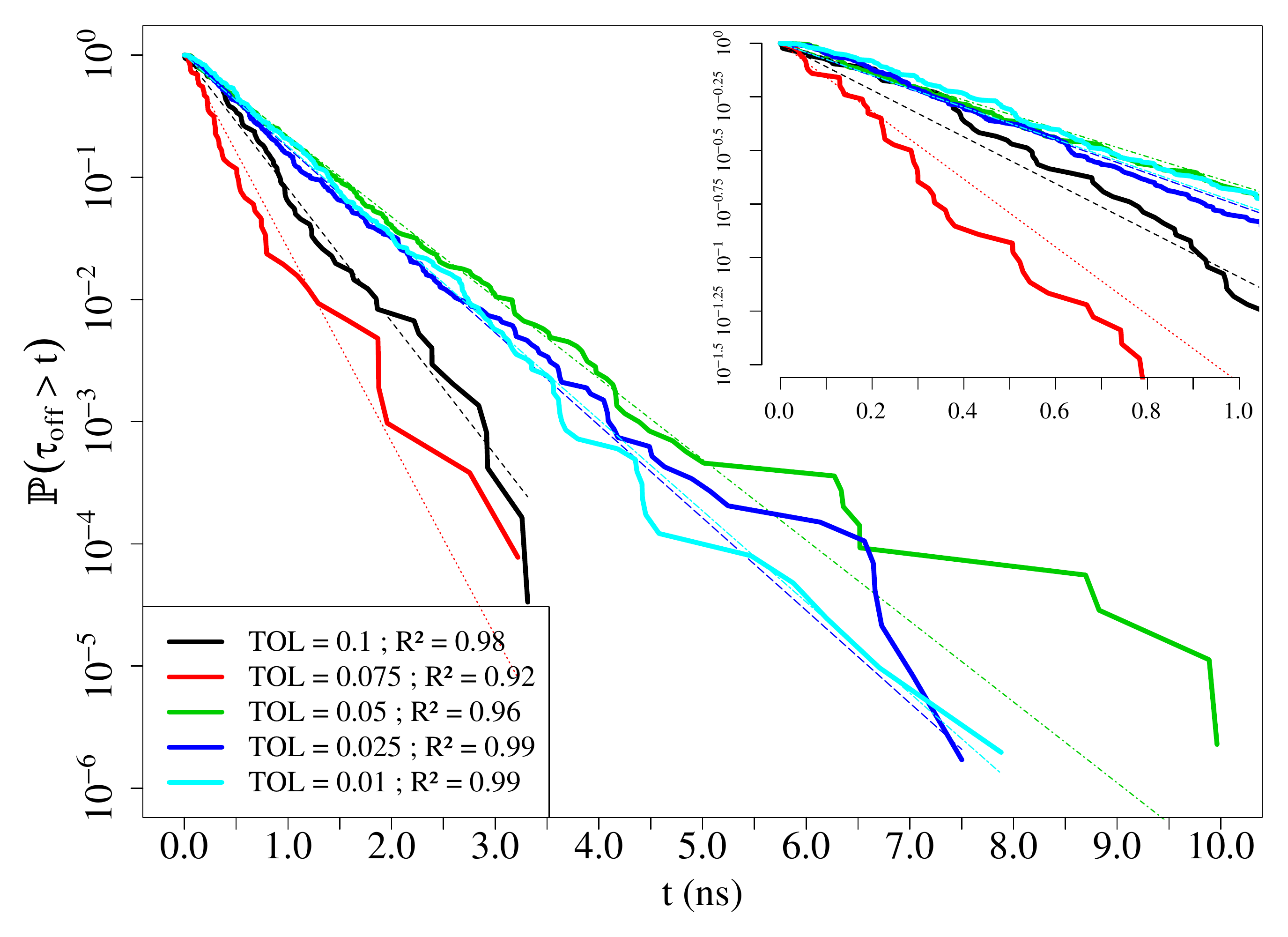}
\caption{Distribution of the dissociation time $\tauoff$ for the complex FKBP--DMSO, 
estimated using the \gparrep method, at different levels of tolerance
$\tol = \{0.1,0.075,0.05,0.025,0.01\}$.
The top-right inset is a zoom for $t < 1.0~\rm ns$, and dashed straight lines (for which the $R^2$ coefficient of determination is given)
denote the expected quasi-exponential distribution for large t: $\ln \mathbb{P}(\tauoff >t) = -t/\mathbb{E}(\tauoff)$. 
See Table~\ref{tab:fkbp_results} for quantitative values of: $\mathbb{E}(\tauoff)$, $N$, $n$ and the $95 \%$ confidence interval.
\label{fig:FKBP_tau_off_distrib}
}
\end{figure}

\begin{figure}[h!]
\centering
\includegraphics[width=0.85\linewidth]{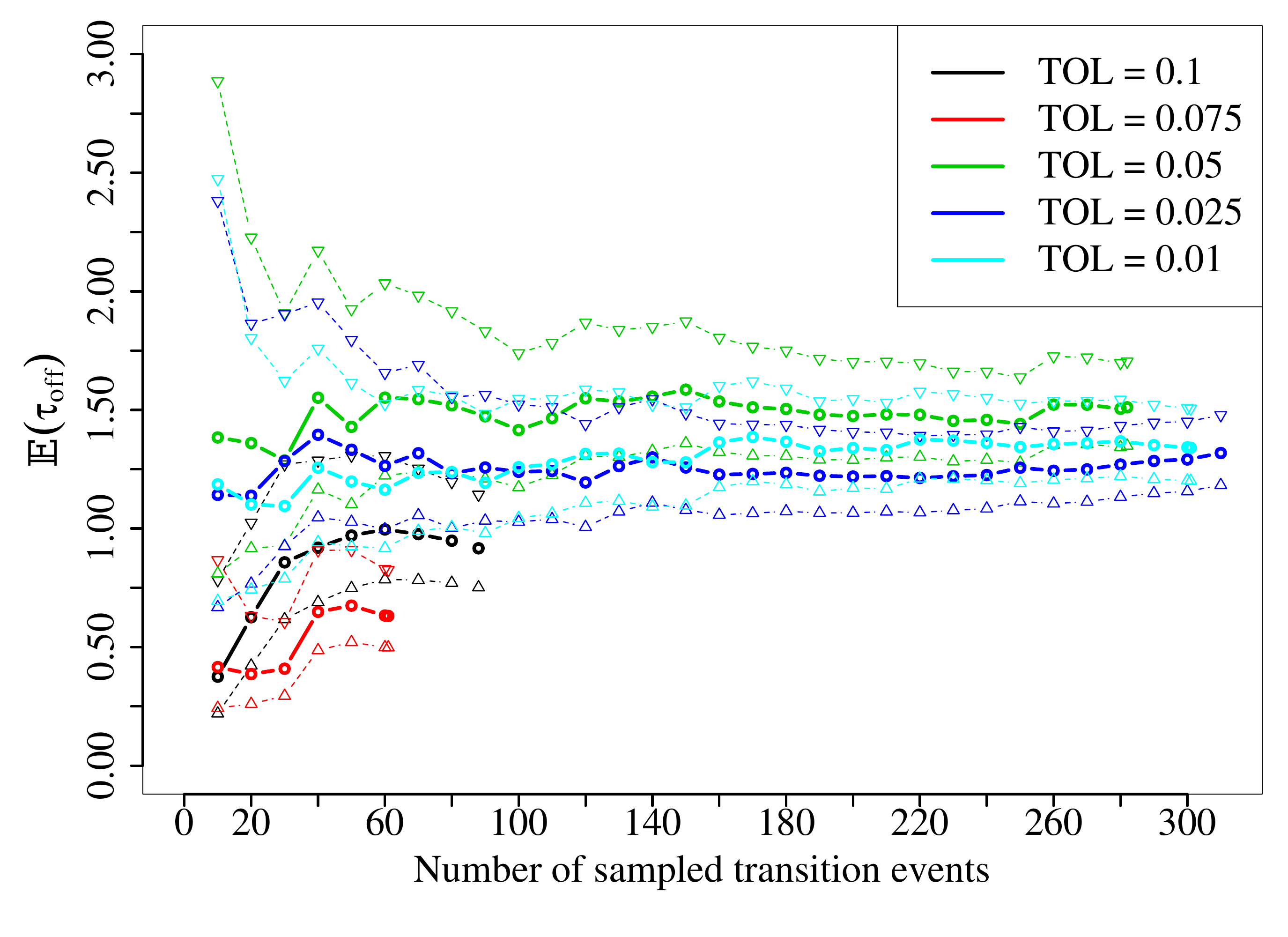}
\caption{
Convergence of $\mathbb{E}(\tauoff)$ for the datasets from Figure~\ref{fig:FKBP_tau_off_distrib}
and Table~\ref{tab:fkbp_results}. Dashed lines correspond to the $95 \%$ confidence interval.
\label{fig:FKBP_tau_off_convergence}
}
\end{figure}

In Figure~\ref{fig:FKBP_tau_off_distrib} the distribution of $\tauoff$ for tolerance values of $\tol = \{0.1,0.075,0.05,0.025,0.01\}$
is shown (details are available in Table~\ref{tab:fkbp_results}):
first, a moderate number of transition events ($n < 100$)
was generated for each level of $\tol$, and a first estimate of $\mathbb{E}(\tauoff)$ was calculated: as the results for $\tol = 
\{0.1,0.075\}$ appeared to be far from the results obtained for more stringent tolerances, they were not further considered;
then for the three remaining tolerance levels ($\tol = \{0.05,0.025,0.01\}$), extended simulations were performed which permitted to obtain 
$n = \{282,320,301\}$ samples of $\tauoff$, thus providing estimates $\mathbb{E}(\tauoff)$ of respectively $1.51$, $1.32$ and $1.34~\rm ns$ 
with a confidence interval of approximately $\pm 0.3~\rm ns$ (see Table~\ref{tab:fkbp_results});
the convergence of the corresponding simulations can be observed on Figure~\ref{fig:FKBP_tau_off_convergence}.

It should first be noted that levels of $\tol$ larger than $0.05$ have to be avoided for this system
(and, from our experience, for any application in general)
as they will systematically produce biased results: it is indeed not 
possible to generate initial conditions distributed according to the QSD by using such a loose tolerance criterion, especially 
when the definition of the state $\Omega$ involves a large number of degrees of freedom.

For tolerance levels smaller or equal to 0.05, the confidence intervals mostly overlap, as one can see on 
Figure~\ref{fig:FKBP_tau_off_convergence}: for $\tol = 0.01$ or $0.025$ (respectively the cyan and dark blue lines) the two 
estimated values of $1.32$ and 
$1.34~\rm ns$ are almost identical,
for $\tol = 0.05$ (the green line)
the estimated value of $\mathbb{E}(\tauoff)$ is $1.51$, a slightly higher value.
A closer look at the upper and lower bounds of the confidence intervals shows a constant overlap --- of decreasing width --- around 
$1.35~\rm ns$, which is the value of $\mathbb{E}(\tauoff)$ for 
$\tol = 0.01$ or $0.025$ ($1.34$ and $1.32~\rm ns$), therefore suggesting that, once again, strict tolerance criteria provide the most 
accurate estimate;
this is confirmed by a qualitative (solid vs dashed lines) and quantitative (coefficient $R^2$) look at 
Figure~\ref{fig:FKBP_tau_off_distrib}, where one can see that the two lowest values of $\tol$ follow the more accurately the 
quasi-exponential distribution (however, one should remember that the distribution is not expected to be exactly exponential especially for 
small values of $\tauoff$, as exit events of the reference walker happening before the end of the Convergence step are not guaranteed to be 
exponentially distributed as the QSD was not yet reached).

In the aforementioned Ref. \cite{xu_fkbp_2016} the estimate of $\mathbb{E}(\tauoff)$ is $2.2~\rm ns$ (no confidence interval provided):
this can be considered to be in a reasonable agreement with our value of $1.34~\rm ns$ obtained for $\tol = 0.01$ and where the $95 \%$ 
confidence interval is $(1.20,1.51)~\rm ns$, considering 
that the force-field was different, and that the current study uses an implicit solvent while the reference used explicit water molecules.\\

\begin{table}[h!]
\centering
\caption{
Summary of the estimated value of $\mathbb{E}(\tauoff)$ and of the corresponding $95 \%$ confidence 
interval for data presented in Figures \ref{fig:FKBP_tau_off_distrib} and \ref{fig:FKBP_tau_off_convergence}.
\label{tab:fkbp_results}
}
\begin{tabular}{c|c|c|c|c}
   $\tol$   &  $N$  &  $n$  & $\mathbb{E}(\tauoff)$ (ns) & confidence interval (ns) \\ \hline
	 $0.1$  & $112$ & $88$  &           $0.92$           &    $(0.75,1.14)$     \\
	$0.075$ & $112$ & $61$  &           $0.63$           &    $(0.50,0.83)$     \\
	$0.05$  & $140$ & $282$ &           $1.51$           &    $(1.35,1.70)$     \\
	$0.025$ & $140$ & $320$ &           $1.32$           &    $(1.19,1.48)$     \\
	$0.01$  & $140$ & $301$ &           $1.34$           &    $(1.20,1.51)$
\end{tabular}
\end{table}

%-----------------------------
\subsubsection{Performance and convergence to the QSD}

Because the FKBP--DMSO system is much more representative of a typical research application than the alanine dipeptide, it is of an utmost 
interest to provide an accurate estimate of the performances:
for this, we provide in Table~\ref{tab:fkbp_benchmark} benchmarking data (following the same methodology established for Table
\ref{tab:ala2_benchmark}): the three datasets correspond to the samples for $\tol = \{0.01,0.025,0.05\}$
from Figure~\ref{fig:FKBP_tau_off_distrib}; the number of replica $N = 140$ corresponds to the maximum speedup, while column 
$5$ reports the calculated effective speedup; the ratio given in column $6$ gives an idea of the efficiency of the implementation for
studying this medium size protein--ligand system.

One can see that the effective speedup is close to $97$ (i.e. $\approx 69 \%$ of the maximum $N = 140$) for tolerances of $0.025$ and 
$0.05$; for the 
stricter tolerance 
level of $0.01$ the speedup falls to $\approx 80$ (i.e. $\approx 57 \%$ of the maximum $N = 140$) which illustrates the computational cost 
one has to 
pay for an increased accuracy.
Once again the ability to obtain a speedup between $57$ and $69~\%$ of the maximum possible denotes the parallel efficiency of the \gparrep
implementation on a production system, and is an extremely promising achievement towards future studies of larger biochemical systems.

\begin{figure}[h!]
\centering
\includegraphics[width=0.85\linewidth]{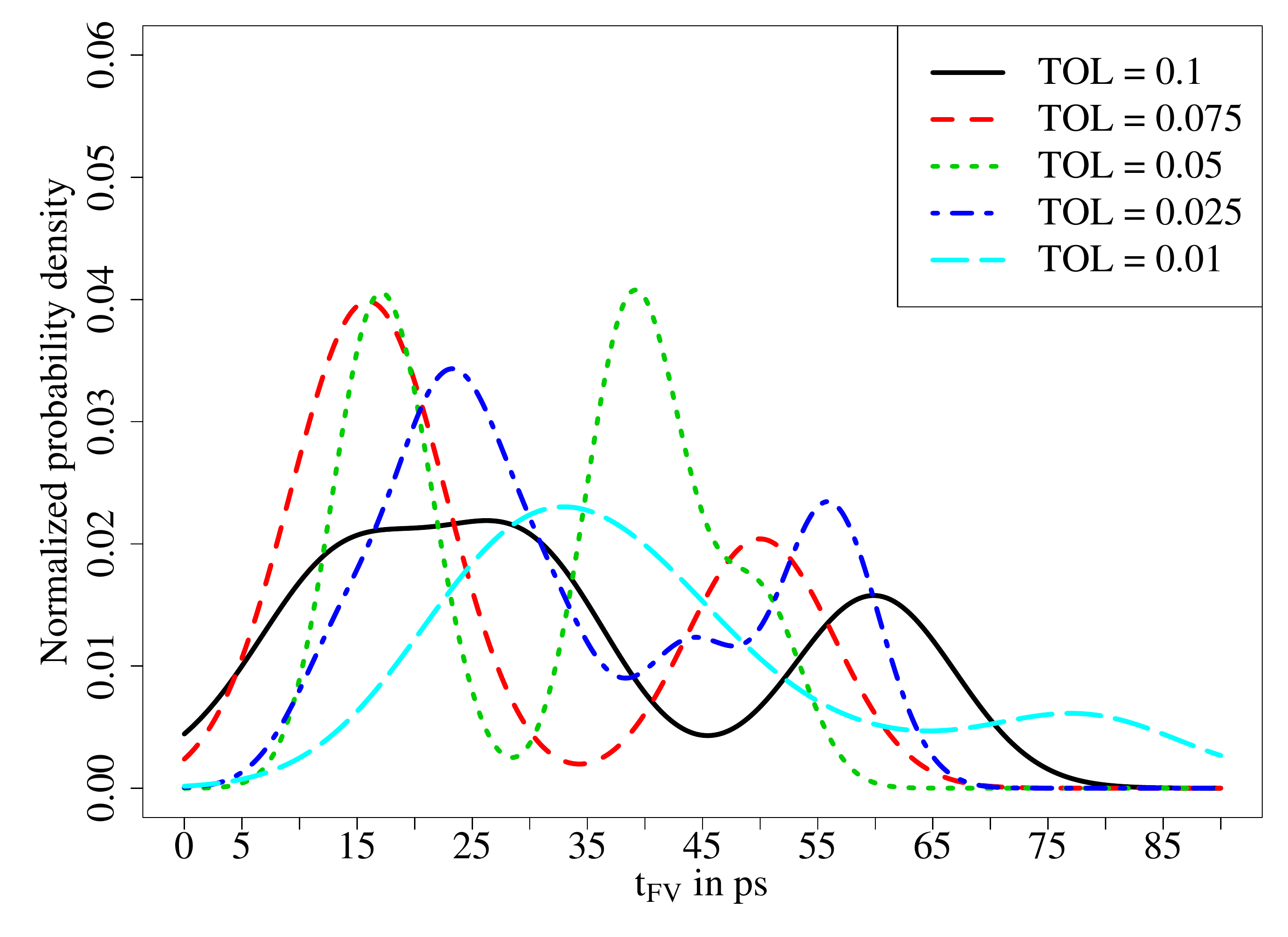}
\caption{
Kernel Density Estimation of the distribution of $\tfv$, obtained
for tolerance levels of $\tol = \{0.1,0.075,0.05,0.025,0.01\}$ (datasets from Figure~\ref{fig:FKBP_tau_off_distrib} and 
Table~\ref{tab:fkbp_results}).
\label{fig:fkbp_tfv_distrib}
}
\end{figure}

The distribution of $\tfv$ is illustrated in Figure~\ref{fig:fkbp_tfv_distrib} for all 
the considered levels of tolerance: all distributions appear to be multimodal, revealing that multiple sub-states 
are likely to 
be found within the surrounding cavity definition of the bound state (this was suggested in earlier studies such as 
Ref. \cite{huang_fkbp_small_2011});
for the larger levels of $\tol$ the multimodality is particularly visible with two well defined peaks, resulting in an average $\tfv$ 
falling 
between the peaks, around $\approx 25~\rm ps$; however for low tolerance such as $0.01$ a broad distribution is observed, and the average 
$\tfv$ goes to $\approx 50~\rm ps$.

This emphasizes once again how difficult it would be to fix a priori a value for $\tfv$ (as required by the original ParRep 
implementations), as this value would be inappropriate for some of the initial conditions; the ability of the \gparrep algorithm to 
automatically determine a value of $\tfv$ appropriate for the current initial condition therefore appears to be a major advantage of the 
method when investigating protein--ligand complexes' dissociation.

\begin{table}[h!]
\centering
\caption{
Benchmarking data for three of the datasets from Figure~\ref{fig:FKBP_tau_off_distrib} and Table~\ref{tab:fkbp_results}
(corresponding to $N = 140$,  $\tol = \{0.01,0.025,0.05\}$): each replica used $P = 4$ CPUs cores,
and the equivalent speed of a reference Langevin dynamics on those same $4$ cores is $5.15$ ns/day;
see Table~\ref{tab:ala2_benchmark} for the methodology.
\label{tab:fkbp_benchmark}
}
\begin{tabular}{l|c|c|c|c|c}
	$\tol$         &  WT(s)  & $\tsim$(ns) & Speed (ns/day) & Eff. speedup & (Eff./Max.) \\ \hline
	$0.01$         & $85142$ &   $403.5$   &    $409.4$     &    $79.5$    &  $56.8 \%$  \\
	$0.025$        & $79574$ &   $457.6$   &    $496.8$     &    $96.5$    &  $68.9 \%$  \\
	$0.05$         & $84455$ &   $482.2$   &    $493.4$     &    $95.8$    &  $68.4 \%$
\end{tabular}
\end{table}

\clearpage
%%%%%%%%%%%%%%%%%%%%%%%%%%%%%%%%%%%%%%%%%%%%%%%%%%%%%%%%%%%%%%%%%%%%%%%%%%%%
\section{Conclusion and outlook}
\label{sec:ccl}

%-----------------------------
% concerning implementation
%-----------------------------

In this article, we detailed a new implementation of the \gparrep algorithm, developed with the aim of facilitating
the study of biochemical systems exhibiting strong metastability.
After detailing the methods and the software implementation in Sections~\ref{sec:methods}~and~\ref{sec:softwareImpl},
a validation (Section \ref{sec:results}) with two systems of increasing complexity was discussed.\\

%-----------------------------
% concerning ala2
%-----------------------------

In subsection~\ref{subsec:ala2}, it was shown that the \gparrep method can accurately sample the transition time $\tauala$ characterizing 
the conformational equilibrium of alanine dipeptide in vacuum: for sufficiently small levels of tolerance (e.g. $\tol = 0.01$), the 
estimation converges 
to what was obtained from a long reference Langevin dynamics (see Table~\ref{tab:ala2_stats} for a summary, and Figures 
\ref{fig:ala2_states_rama} to \ref{fig:ala2_tfv_distrib}).
Results also appeared to compare favorably to previously published studies \cite{ala2_Strodel_2008}.
Finally it was also shown that the implementation of the algorithm proves to be scalable as one can obtain $\approx 80 \%$ of the maximum 
possible speedup (see Table~\ref{tab:ala2_benchmark}).
\\

%-----------------------------
% concerning kkbp--dmso
%-----------------------------

The second application consisted in the study of the dissociation of the FKBP--DMSO complex (subsection~\ref{subsec:fkbp}), a 
protein--ligand system of larger size, much more representative of typical metastable problems encountered in computational biology or 
chemistry.
The goal was to obtain an accurate estimate of the average time $\mathbb{E}(\tauoff)$ required for observing a dissociation of the complex,
with comparison to previous computational studies \cite{huang_fkbp_free_2011,huang_fkbp_small_2011,xu_fkbp_2016}.
It was shown (see Table~\ref{tab:fkbp_results} and the associated Figures \ref{fig:FKBP_b_u} to \ref{fig:fkbp_tfv_distrib})
that a simple definition of the bound and unbound states based on a two distances threshold can provide an accurate estimate of 
$\mathbb{E}(\tauoff)$, once again when tolerance levels $\tol < 0.05$ are used: a value of $1.32 < \mathbb{E}(\tauoff) < 1.51$ is found 
using the \gparrep method,
the value of $1.34~\rm ns$ for $\tol = 0.01$ appearing to be the most accurate, and this compares relatively well to Ref. 
\cite{xu_fkbp_2016}
where a value of $2.2~\rm ns$ was found using a different force-field and an explicit solvent.
The algorithm was also benchmarked for the FKBP--DMSO system (see Table~\ref{tab:fkbp_benchmark}), and it was shown
that one can maintain performances up to $60 \sim 70 \%$ of the maximum possible speedup on 560 CPU cores,
which definitely makes this new \gparrep ready for production on large scale HPC machines.\\

%-----------------------------
% concerning Work in progress
%-----------------------------

From the two studies performed in this article it ought to be remembered that, beyond the accurate definition of the states $\mathcal{S}$, 
the choice of the tolerance level is the main parameter
influencing the accuracy of the results:
a value of $\tol = 0.01$ appears to be the most reasonable choice, confirming previous observations on smaller systems 
\cite{binder_genparrep_2015}.
Furthermore, the algorithm provides an accurate estimate of the time $\tfv$ required for approximating the QSD depending on the initial 
condition within the state, and it was shown 
(see Figures \ref{fig:ala2_tfv_distrib} and \ref{fig:fkbp_tfv_distrib}) that $\tfv$ is distributed over a large interval of time: in 
such a case the a priori choice of a fixed value decorrelation time approximating $\tfv$ (as it was done in the original ParRep 
implementations)  is non-obvious, and the use of the \gparrep method is justified.\\

As an outlook, it has to be emphasized that there is still, of course, place for improvement of the software implementation:
the authors would like to make the program compatible with other MD engines;
tests are currently being performed where replicas are distributed over General-Purpose computing units (GPGPUs) in order to consider 
applications to larger systems;
and preliminary simulations are being performed on a HPC Cloud Computing platform, on which a user could easily use thousands of 
replicas. 

The authors are also currently studying more 
advanced biochemical metastable problems, including larger protein--ligand systems in explicit water where the states consist in a set of 
disjoint cavities inside a protein.
\\

%%%%%%%%%%%%%%%%%%%%%%%%%%%%%%%%%%%%%%%%%%%%%%%%%%%%%%%%%%%%%%%%%%%%%%%%%%%%%%%%%%%%%%%%%%%%%%%%%%%%%%%%%%%%%%%%%%%%%%%
% ARTICLE ENDS HERE
%%%%%%%%%%%%%%%%%%%%%%%%%%%%%%%%%%%%%%%%%%%%%%%%%%%%%%%%%%%%%%%%%%%%%%%%%%%%%%%%%%%%%%%%%%%%%%%%%%%%%%%%%%%%%%%%%%%%%%%

\section*{Acknowledgement}

\begin{itemize}
\item This work is supported by the European Research Council under the European Union's Seventh Framework Program
(FP/2007-2013)/ERC Grant Agreement number 614492.

\item The authors acknowledge the \href{http://www.maisondelasimulation.fr/}{``Maison de la Simulation''}
(and particularly Matthieu Haefele and Julien Derouillat) for helpful discussion and advices concerning the optimization of the software.

\item This work was granted access to the HPC resources of the CINES under the
GENCI allocations 2017-AP010710193, 2017-AP010710245 and 2017-A0030710275.

\item FH acknowledges the Institute for Pure and Applied Mathematics (IPAM) at the University of California Los Angeles (UCLA)
for participation to the long program ``Complex High-Dimensional Energy Landscapes'', its organizers, participants, and particularly 
members of the ``Accelerated dynamics and benchmarking'' working group for discussion and remarks on this work.

\end{itemize}

\section*{Bibliography}

\bibliographystyle{elsarticle-num} 
\bibliography{parrep}

\end{document}